\documentclass[12pt]{JHEP3}
\usepackage{epsfig,latexsym,amsmath,amssymb,bm,cite}
\usepackage{graphicx}

\newcommand{\be}{\begin{equation}}
\newcommand{\ee}{\end{equation}}
\newcommand{\bea}{\begin{eqnarray}}
\newcommand{\eea}{\end{eqnarray}}
\newcommand{\bem}{\begin{multline}}
\newcommand{\eem}{\end{multline}}
\newcommand{\beg}{\begin{gather}}
\newcommand{\eeg}{\end{gather}}

\newcommand{\stackeven}[2]{{{}_{\displaystyle{#1}}\atop\displaystyle{#2}}}

\newcommand{\gsim}{\stackeven{>}{\sim}}
\newcommand{\as}{\alpha_s}

\def\eq#1{{Eq.~(\ref{#1})}}
\def\fig#1{{Fig.~\ref{#1}}}
\newcommand{\ben}{\begin{eqnarray*}}
\newcommand{\een}{\end{eqnarray*}}

\title{Modeling Heavy Ion Collisions in AdS/CFT}

\author{ Javier L.\ Albacete, \ Yuri V.\ Kovchegov, \ Anastasios Taliotis
\\~~\\ Department of Physics, The Ohio State University,
Columbus, OH 43210,USA \\~~\\ E-mail addresses: \email{albacete@mps.ohio-state.edu}, 
\email{yuri@mps.ohio-state.edu}, \email{taliotis.1@osu.edu}}

\date{May 2008}

\abstract{We construct a model of high energy heavy ion collisions as
  two ultrarelativistic shock waves colliding in AdS$_5$. We point out
  that shock waves corresponding to physical energy-momentum tensors
  of the nuclei completely stop almost immediately after the collision
  in AdS$_5$, which, on the field theory side, corresponds to complete
  nuclear stopping due to strong coupling effects, likely leading to
  Landau hydrodynamics. Since in real-life heavy ion collisions the
  large Bjorken $x$ part of nuclear wave functions continues to move
  along the light cone trajectories of the incoming nuclei leaving the
  small-$x$ partons behind, we conclude that a pure large coupling
  approach is not likely to adequately model nuclear collisions.  We
  show that to account for small-coupling effects one can model the
  colliding nuclei by two (unphysical) ultrarelativistic shock waves
  with zero net energy each (but with non-zero energy density). We use
  this model to study the energy density of the strongly-coupled
  matter created immediately after the collision.  We argue that
  expansion of the energy density $\epsilon$ in the powers of proper
  time $\tau$ squared corresponds on the gravity side to a
  perturbative expansion of the metric in graviton exchanges.  Using
  such expansion we reproduce our earlier result
  \cite{Kovchegov:2007pq} that the energy density of produced matter
  at mid-rapidity starts out as a constant (of time) in heavy ion
  collisions at large coupling. }


\preprint{}

\begin{document}


\section{Introduction}

There is a mounting phenomenological evidence coming from RHIC data
for a strongly coupled medium created in heavy ion collisions
\cite{Kolb:2000sd,Kolb:2000fh,Huovinen:2001cy,Kolb:2001qz,Heinz:2001xi,Teaney:1999gr,Teaney:2000cw,Teaney:2001av,Teaney:2003kp,Policastro:2001yc,Son:2002sd,Policastro:2002se,Kovtun:2003wp,Kovtun:2004de,Moore:2004tg,CasalderreySolana:2004qm,CasalderreySolana:2006sq,Shuryak:2006se}.
Ideal hydrodynamics simulations have been extremely successful in
describing data generated in heavy ion collisions at RHIC
\cite{Kolb:2000sd,Kolb:2000fh,Huovinen:2001cy,Kolb:2001qz,Heinz:2001xi,Teaney:1999gr,Teaney:2000cw,Teaney:2001av}.
These analyses require very small shear viscosity
\cite{Teaney:2003kp}, indicating that the medium is strongly coupled
\cite{Policastro:2001yc,Son:2002sd,Policastro:2002se,Kovtun:2003wp,Kovtun:2004de},
and a very short thermalization time of the initially produced system,
of the order of $0.3 \div 0.6$~fm/c
\cite{Kolb:2000sd,Kolb:2000fh,Huovinen:2001cy,Kolb:2001qz,Heinz:2001xi,Teaney:1999gr,Teaney:2000cw,Teaney:2001av}.
At the same time, many bulk features of heavy ion collisions at RHIC
which are sensitive to the initial-time dynamics, such as the energy,
rapidity and centrality dependence of particle production are very
well-described in the weakly-coupled framework of the Color Glass
Condensate (CGC)
\cite{Blaizot:1987nc,McLerran:1993ni,McLerran:1993ka,McLerran:1994vd,Kovchegov:1996ty,Kovchegov:1997pc,Kovner:1995ja,Kovchegov:1997ke,Krasnitz:1998ns,Krasnitz:1999wc,Krasnitz:2003jw,Kovchegov:2000hz,Lappi:2003bi,Kharzeev:2000ph,Kharzeev:2001yq,Kharzeev:2002pc,Kharzeev:2004yx,Albacete:2003iq,Albacete:2007sm}
(for a review of CGC see
\cite{Iancu:2003xm,Weigert:2005us,Jalilian-Marian:2005jf}). In the CGC
approach a heavy ion collision releases the small Bjorken-$x$ partons
in the nuclear wave functions, which quickly go on mass shell and
become real: the resulting particle distribution in highly anisotropic
in momentum space \cite{Krasnitz:2002mn,Kovchegov:2005ss}. A question
then arises about how this initially weakly-coupled highly anisotropic
system becomes isotropic and equilibrates very quickly, becoming a
strongly-coupled, possibly thermal, medium.

So far, conventional perturbative descriptions of thermalization of
the produced medium
\cite{Baier:2000sb,Arnold:2003rq,Mrowczynski:1988dz,Mrowczynski:1993qm}
have not been able to account for the short thermalization time of the
order of a fraction of a fermi required by hydrodynamic simulations to
describe the data.  Perturbative thermalization in heavy ion
collisions was also studied in
\cite{Kovchegov:2005ss,Kovchegov:2005kn,Kovchegov:2005az}, where it
was concluded that perturbative thermalization scenarios are not
likely to be compatible with RHIC data. This leads one to conclude
that it is highly likely that non-perturbative large-coupling QCD
effects are responsible for the apparent thermalization observed in
RHIC data.  While the research on perturbative thermalization
scenarios along the lines outlined in
\cite{Baier:2000sb,Arnold:2003rq,Mrowczynski:1988dz,Mrowczynski:1993qm}
is vigorously pursued in the community, we believe it probable that
the dynamics of the medium produced in a heavy ion collision at RHIC
proceeds as follows. The medium starts out being weakly-coupled, being
well-described by CGC. After a very short proper time, of the order of
$\tau \gsim 1/Q_s \approx 0.2 \div 0.3$~fm/c (and possibly at $\tau
\approx 1/\Lambda_{\text{QCD}} \approx 1$~fm/c) the coupling becomes
strong. Strong coupling effects are likely to quickly thermalize the
medium, allowing for its hydrodynamic description.

Unfortunately to date there is no consistent way to describe both the
weakly-coupled and the strongly coupled dynamics of QCD medium in a
unified framework. However, as we argued above, to understand the
general physical nature of thermalization (and, more importantly,
isotropization) of the medium, and to see whether isotropization and
thermalization take place at all, a purely strong coupling approach
seems, a priori, appropriate. Indeed at strong coupling QCD becomes
non-perturbative and no controllable dynamical calculation appears to
be possible. Instead one could use the Anti-de Sitter space/conformal
field theory (AdS/CFT) correspondence
\cite{Maldacena:1997re,Gubser:1998bc,Witten:1998qj,Aharony:1999ti} to
understand the same process for ${\cal N} =4$ super-Yang-Mills theory.
This correspondence, and in particular the gauge-gravity duality which
follows from it, allows one to understand strong coupling effects in
such QCD-like theories as ${\cal N} =4$ super-Yang-Mills theory using
super-gravity in 5 dimensions.  AdS/CFT correspondence has been useful
in providing insight in the behavior of the shear viscosity in
strongly-coupled gauge theories
\cite{Policastro:2001yc,Son:2002sd,Policastro:2002se,Kovtun:2003wp,Kovtun:2004de}
along with providing other interesting results on the evolution of the
medium created in heavy ion collisions
\cite{Janik:2005zt,Janik:2006gp,Janik:2006ft,Nakamura:2006ih,Shuryak:2005ia,Lin:2006rf,Lublinsky:2007mm,Nastase:2005rp,Bak:2006dn,Kajantie:2006hv,Kajantie:2006ya,Kajantie:2007bn,Kovchegov:2007pq,Kajantie:2008rx,Grumiller:2008va}.

The goal of this paper is to make progress in constructing a dual
geometry in AdS$_5$ space for heavy ion collisions with the goal of
understanding the onset of isotropization and thermalization of the
produced medium. The previous paper on the subject by two of the
authors \cite{Kovchegov:2007pq} studied the very early time dynamics
of the medium produced in the collisions.  It was shown that, assuming
rapidity-independence of the produced medium and assuming
non-negativity of its energy density, one would obtain that the energy
density of the produced medium should start out as a constant of time
at very early times immediately after the collisions. This implied
that the longitudinal pressure of this early-time medium is negative
and the medium is thus highly anisotropic. This behavior is similar to
that of the weakly-coupled CGC medium at early times
\cite{Lappi:2003bi,Fukushima:2007ja,Fries:2006pv}. The problem of
isotropization and the onset of Bjorken hydrodynamics
\cite{Bjorken:1982qr} in this framework can be formulated as the
question about understanding the transition from the negative
longitudinal pressure of the medium at early times to the positive
longitudinal pressure (comparable to the transverse pressure) at late
times \cite{Kovchegov:2005ss}. In
\cite{Janik:2005zt,Janik:2006ft,Heller:2007qt} it was shown that the
dynamics of a strongly-coupled rapidity-independent medium leads to
Bjorken hydrodynamics behavior at late proper times: it is therefore
likely that isotropization transition takes place at some intermediate
time between the early-time dynamics of \cite{Kovchegov:2007pq} and
the late time dynamics of
\cite{Janik:2005zt,Janik:2006ft,Heller:2007qt}.

To better understand this transition one needs to find the
energy-momentum tensor of the medium at a later times than considered
in \cite{Kovchegov:2007pq}. Unfortunately, a consistent expansion of
the energy-momentum tensor in the powers of proper time $\tau$
requires some knowledge of the geometry dual to the colliding nuclei.
(In \cite{Kovchegov:2007pq} nothing was assumed about the colliding
nuclei, except that they lead to rapidity-independent distribution of
matter.) Thus in this paper we try to construct a geometry dual to the
collision of the two nuclei. We model two nuclei as shock waves in
AdS$_5$. Modeling nuclei with shock waves has previously been
considered in \cite{Kajantie:2008rx} for AdS$_3$ corresponding to
gauge theory in two space-time dimensions and in
\cite{Nastase:2005rp,Grumiller:2008va} for AdS$_5$.

The paper is organized as follows. We begin by spelling out some
general formulas used in the paper in Sect. \ref{general}. We proceed
by setting up the problem in Sect. \ref{setup}. We start with two
colliding shock waves, each given by the metric like that shown in
\eq{1nuc} (see \cite{Janik:2005zt}). We argue that this metric
corresponds to a single graviton field, as shown in \fig{1gr}. We then
argue that a consistent expansion of the metric at the time after the
collision can be constructed by considering higher order graviton
exchanges between the boundary and the bulk, as shown in \fig{pert}.
We construct a general next-to-leading order perturbative contribution
to the metric in graviton exchanges in Sect.  \ref{gensol}. The
solution is given by Eqs. (\ref{metric2}), (\ref{h0int}),
(\ref{h1int}), (\ref{hsol}), (\ref{gsol}), (\ref{fsol}), (\ref{lam}),
(\ref{fdsol}), (\ref{lam2}).

Using the obtained solution we study the collision of two physical
shock waves in Sect. \ref{real_shock}. We conclude that the shock
waves, and the nuclei in the boundary theory, completely stop very
shortly after the collision, after a time of the order of inverse
typical transverse momentum scale in the problem, as given in
\eq{stoptime2}. We interpret this result as creation of a black hole,
similar to what is suggested for collisions of particles at
transplanckian energies
\cite{Giddings:2001bu,Eardley:2002re,Kancheli:2002nw}, though our
black hole would be created in the bulk. On the gauge theory side this
implies that strong coupling effects would completely stop the nuclei
shortly after the collision, on the time scale less than or equal to
$1$~fm/c. Such strong coupling effects are likely to thermalize the
system soon after the stopping, leading to Landau hydrodynamic
description of the system \cite{Landau:1953gs}. This thermalization
scenario is very different from the onset of Bjorken hydrodynamics
outlined above. It is possible in principle and has been advocated in
the literature \cite{Steinberg:2004vy}.

However, we believe that the Landau hydrodynamic scenario is not
likely to be relevant for heavy ion collisions. This claim is
supported by the following observations. On the one hand, the ideal
Bjorken hydrodynamics has been extremely successful in describing the
particle spectra and elliptic flow
\cite{Kolb:2000sd,Kolb:2000fh,Huovinen:2001cy,Kolb:2001qz,Heinz:2001xi,Teaney:1999gr,Teaney:2000cw,Teaney:2001av}.
On the other hand, the agreement with experimental data of the
predictions based on the weakly-coupled CGC approaches to description
of particle multiplicities in nuclear collisions
\cite{Kharzeev:2000ph,Kharzeev:2001yq,Kharzeev:2002pc} indicates that
the complete nuclear stopping does not happen in the actual
collisions. Indeed, in high-energy scattering at small coupling the
hard (large Bjorken-$x$) parts of the nuclear wave functions simply go
through each other without recoil. In the quasi-classical CGC limit
this leads to rapidity-independent, Bjorken-like picture of particle
production in heavy ion collisions
\cite{Kovner:1995ja,Kovner:1995ts,Kovchegov:1997ke,Gyulassy:1997vt,Krasnitz:2003nv}.
Thus it appears that weak coupling effects are a {\sl key} ingredient
for a proper description of the space-time structure of heavy ion
collisions, even in the medium becomes strongly coupled shortly after
the collision.

As we do not know how to model weak coupling effects in AdS/CFT, we
try in Sect.  \ref{our_shock} to mimic them by introducing unphysical
shock waves with non-positive-definite energy density.  Such shock
waves are indeed unphysical and can not follow from the underlying
string theory. Even for the gravity in the bulk these shock waves can
only serve as sources external to the theory. An example of the
energy-momentum tensor of such shock wave is given in \eq{Tzero}.
Using the general solution of Sect. \ref{gensol}, we construct the
early time energy density and pressures of the medium produced in the
``collision'' of two of such shock waves, which are shown in
\eq{epp3}. As can be seen from \eq{epp3}, the energy density starts
out as a constant in time. We have thus reproduced the results of
\cite{Kovchegov:2007pq}, but now in a more dynamical setting. The
problem of thermalization formulated with the help of these unphysical
shock waves is probably that of isotropization of the produced medium
to achieve Bjorken hydrodynamics \cite{Bjorken:1982qr}, as described
above and considered in \cite{Kovchegov:2005ss,Kovchegov:2007pq}.

We discuss possible higher order corrections to the result of
\eq{epp3} in Sect. \ref{dilaton} and observe that a dilaton field may
need to be introduced at higher orders to account for initial
non-equilibrium between chromo-electric and chromo-magnetic modes in
the medium (see \eq{dile}). We conclude in Sect. \ref{conc} by
restating our main results.

Our solution found in Sect. \ref{gensol} is general and is valid for
any shock wave profile, unlike the solution found in
\cite{Grumiller:2008va} which works only for delta-function shock
waves. In the particular case of the delta-function shock waves, our
solution reduces to that found in \cite{Grumiller:2008va}. The general
nature of our solution allowed us in Sect.  \ref{real_shock} to reach
a more general physical conclusion that the collision of any two
physical shock waves (with positive definite energy density) leads to
complete stopping of the shock waves after the collision, probably
leading to the formation of a black hole. In particular this
conclusion applies to the delta-function shock waves used in
\cite{Grumiller:2008va}. Our solution also allows us to tackle
unphysical shock waves in Sect.  \ref{our_shock}, which are impossible
to handle in the formalism of \cite{Grumiller:2008va}. We should also
note that the stopping of physical shock waves found here does not
happen in AdS$_3$ (see \cite{Kajantie:2008rx}), since in 1+1
dimensional gauge theory the nuclei are point particles and stopping
for them is impossible, as there is no transverse directions in which
the momentum could be channeled.


\section{Some Generalities}
\label{general}

Throughout this paper we will work with the metric of AdS$_5$ written
in terms of Fefferman-Graham coordinates \cite{F-G}
\begin{align}
  ds^2 \, = \, \frac{L^2}{z^2} \, \left\{ {\tilde g}_{\mu\nu} (x, z)
    \, d x^\mu \, d x^\nu + d z^2 \right\}
\end{align}
where $\mu, \nu$ run from $0$ to $3$ and $z$ is the coordinate
describing the 5th dimension. The boundary of the AdS space is at $z=0$
and $L$ is the curvature radius of the AdS space.

According to holographic renormalization \cite{deHaro:2000xn}, if one
expands the 4-dimensional metric ${\tilde g}_{\mu\nu} (x, z)$ near the
boundary of the AdS space
\begin{align}
  {\tilde g}_{\mu\nu} (x, z) \, = \, {\tilde g}_{\mu\nu}^{(0)} (x) + z^2 \,
  {\tilde g}_{\mu\nu}^{(2)} (x) + z^4 \,
  {\tilde g}_{\mu\nu}^{(4)} (x) + \ldots , 
\end{align}
then, for Minkowski metric $ {\tilde g}_{\mu\nu}^{(0)} (x) =
\eta_{\mu\nu}$, one gets ${\tilde g}_{\mu\nu}^{(2)} (x) = 0$ and the
expectation value of the energy-momentum tensor of the gauge theory is
\begin{align}\label{holo}
  \langle T_{\mu\nu} \rangle \, = \, \frac{N_c^2}{2 \, \pi^2} \,
  {\tilde g}_{\mu\nu}^{(4)} (x).
\end{align}

Below we will use the light cone coordinates 
\begin{align}
  x^\pm = \frac{x^0 \pm x^3}{\sqrt{2}}
\end{align}
where $x^3$ is the collision axis of the two colliding nuclei.  In
these coordinates the empty AdS$_5$ metric is
\begin{align}
  ds^2 \, = \, \frac{L^2}{z^2} \, \left\{ -2 \, dx^+ \, dx^- + d
    x_\perp^2 + d z^2 \right\},
\end{align}
where $d x_\perp^2 = (d x^1 )^2 + (d x^2)^2$ with $x^1$ and $x^2$ the
transverse dimensions which we will denote using Latin indices, e.g.
$x^i$.  To describe nuclear collisions we will also use the proper
time
\begin{align}
  \tau \, = \, \sqrt{2 \, x^+ \, x^-}
\end{align}
and space-time rapidity
\begin{align}
  \eta \, = \, \frac{1}{2} \, \ln \frac{x^+}{x^-}. 
\end{align}

Einstein equations in AdS$_5$ are
\begin{align}\label{ein1}
  R_{\mu\nu} - \frac{1}{2} \, g_{\mu\nu} \, R + \Lambda_c \,
  g_{\mu\nu} = 0
\end{align}
where $g_{\mu\nu}$ is the full 5-dimensional metric of the AdS$_5$
space, $R$ is the scalar curvature and $\Lambda_c$ is the cosmological
constant. For AdS$_5$ we have
\begin{align}\label{lamc}
  \Lambda_c \, = \, - \frac{6}{L^2}.
\end{align}
and \eq{ein1} gives
\begin{align}\label{curv}
  R \, = \, - \frac{20}{L^2}.
\end{align}
Eqs. (\ref{lamc}), (\ref{curv}) yield
\begin{align}\label{ein2}
  R_{\mu\nu} + \frac{4}{L^2} \, g_{\mu\nu} = 0.
\end{align}

Later in the paper we will also discuss the dynamics of a dilaton
field $\varphi$ coupled to gravity in AdS$_5$. In the presence of
dilaton \eq{ein1} is modified to (see e.g.
\cite{Klebanov:2000me,Balasubramanian:1998de,Heller:2007qt})
\begin{align}\label{ein3}
  R_{\mu\nu} - \frac{1}{2} \, g_{\mu\nu} \, R + \Lambda_c \,
  g_{\mu\nu} \, = \, \frac{1}{2} \, \partial_\mu \varphi \,
  \partial_\nu \varphi - \frac{1}{4} \, g_{\mu\nu} \, \partial_\rho
  \varphi \, \partial^\rho \varphi.
\end{align}
\eq{ein3} can be simplified to give
\begin{align}\label{ein4}
  R_{\mu\nu} + \frac{4}{L^2} \, g_{\mu\nu} \, = \, \frac{1}{2} \,
  \partial_\mu \varphi \, \partial_\nu \varphi. 
\end{align}
The dilaton itself obeys the Klein-Gordon equation in curved
space-time
\begin{align}\label{KG}
  \partial_\mu \left[ \sqrt{-g} \, g^{\mu \nu} \, \partial_\nu \,
    \varphi \right] \, = \, 0.
\end{align}


\section{Setting up the Problem}
\label{setup}

Out goal is to construct a metric in AdS$_5$ which is dual to an
ultrarelativistic heavy ion collision as pictured in \fig{spacetime}.
Throughout the discussion we will use Bjorken approximation of the
nuclei having an infinite transverse extent \cite{Bjorken:1982qr} and
being homogeneous (on the average) in the transverse direction, such
that nothing in our problem would depend on the transverse
coordinates $x^1$, $x^2$.

\FIGURE{\includegraphics[width=10.2cm]{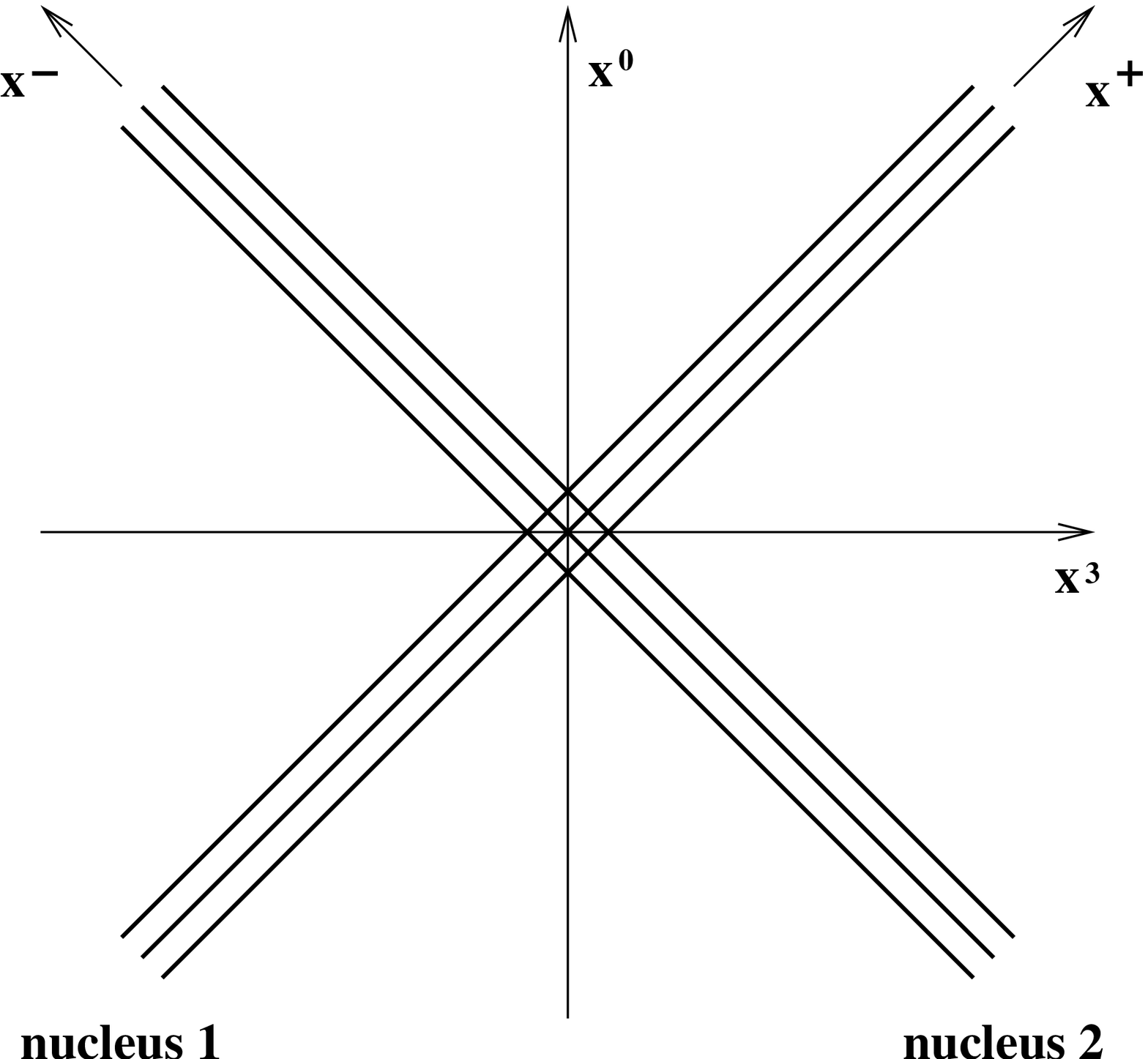}
  \caption{The space-time picture of the ultrarelativistic heavy ion 
    collision in the center-of-mass frame. The collision axis is
    labeled $x^3$, the time is $x^0$.}
  \label{spacetime}
}

We start with a metric for a single ultrarelativistic nucleus moving
along a light cone. As was noted by Janik and Peschanski
\cite{Janik:2005zt} the following metric gives a solution of Einstein
equations in AdS$_5$ in Fefferman-Graham coordinates \cite{F-G} 
\begin{align}\label{1nuc}
  ds^2 \, = \, \frac{L^2}{z^2} \, \left\{ -2 \, dx^+ \, dx^- + \frac{2
      \, \pi^2}{N_c^2} \, \langle T_{--} (x^-) \rangle \, z^4 \, d
    x^{- \, 2} + d x_\perp^2 + d z^2 \right\}.
\end{align}
\eq{1nuc} is a solution of Einstein equations (\ref{ein2}) for any
expectation value of the energy-momentum tensor of the nucleus in four
dimensions $\langle T_{--} (x^-) \rangle$, as long as it is a function
of $x^-$ only \cite{Janik:2005zt}. The factor of $2 \, \pi^2 / N_c^2$
is due to Newton's constant \cite{deHaro:2000xn}. For an
ultrarelativistic nucleus with infinite transverse extent moving along
the $x^+$ axis (see \fig{spacetime}) the leading components of the
energy momentum tensor depend only on $x^-$. Hence the metric in
\eq{1nuc} adequately describes such a nucleus, though does not
restrict the dependence of $\langle T_{--} (x^-) \rangle$ on $x^-$.

While the metric (\ref{1nuc}) is an exact solution of the non-linear
Einstein equations (\ref{ein2}), it can also be represented
perturbatively as a single graviton exchange between the source
nucleus at the AdS boundary and the location in the bulk where we
measure the metric/graviton field. This is shown in \fig{1gr}, where
the solid line represents the nucleus and the wavy line is the
graviton propagator.  Incidentally a single graviton exchange, while
being a first-order perturbation of the empty AdS space, is also an
exact solution of Einstein equations. This means higher order
tree-level graviton diagrams are zero. It is interesting to note that
a similar property has been observed for gauge theories in covariant
gauge \cite{Kovchegov:1996ty,Jalilian-Marian:1997xn}, where the exact
solution of Yang-Mills equations with a single ultrarelativistic
nucleus as a source is given by a single gluon exchange. 

\FIGURE{\includegraphics[width=5cm]{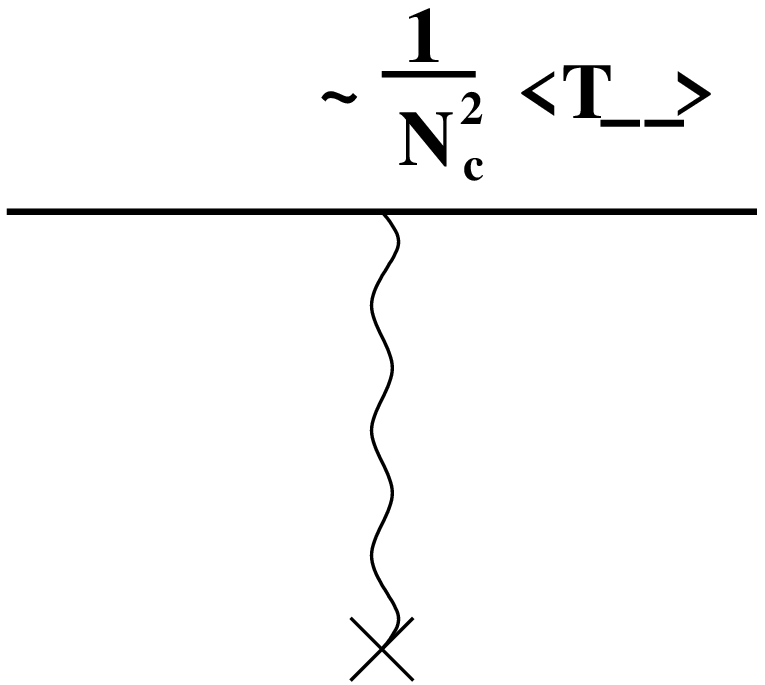}
  \caption{A representation of the metric (\protect\ref{1nuc}) as a graviton 
    (wavy line) exchange between the nucleus at the boundary of AdS
    space (the solid line) and the point in the bulk where the metric
    is measured (denoted by a cross). }
  \label{1gr}
}

As one can see comparing the metric (\ref{1nuc}) with the diagram in
\fig{1gr}, each graviton-nucleus vertex gives a factor
\begin{align}\label{vertex}
  \sim \, \frac{1}{N_c^2} \, \langle T_{--} (x^-) \rangle.
\end{align}
If the nuclear energy-momentum tensor is $N_c$-independent, then in
the large-$N_c$ limit the factor in \eq{vertex} would be small and one
could envision perturbative expansion in this parameter for the
problem of collision of two nuclei. On the other hand, gauge-gravity
duality is valid only in the large-$N_c$ limit: hence we need $\langle
T_{--} (x^-) \rangle \sim N_c^2$ to avoid having $N_c$-suppression for
higher-order graviton exchanges. This could be achieved by imagining a
nucleus with nucleons made out of $N_c^2 -1$ ``valence'' gluons each.
Then $\langle T_{--} (x^-) \rangle \sim N_c^2$ and multiple graviton
exchanges will not be $N_c$-suppressed.

However, one may then worry that the expansion parameter also ceases
to be small. Nevertheless it makes sense to expand in powers of
$\langle T_{--} (x^-) \rangle$, as usually $\langle T_{--} (x^-)
\rangle$ contains some momentum scale characterizing the nucleus such
that one can keep track of the powers of this scale. Hence even if the
expansion parameter is not small, the expansion is still well-defined
and can be kept track of.

For instance, in the original proposal of Janik and Peschanski
\cite{Janik:2005zt}, the energy-momentum tensor due to valence quarks
in the ultrarelativistic nucleus was taken to be
\begin{align}\label{Tmu}
  \langle T_{--} (x^-) \rangle \, = \, \mu \, N_c^2 \, \delta (x^-)
\end{align}
with $\mu$ a scale having dimensions of mass cubed. Starting with the
energy-momentum tensor of a single ultrarelativistic particle and
performing the averaging along the lines shown in Appendix \ref{A}
gives
\begin{align}\label{mu}
  \mu \propto p^+ \, \Lambda^2 \, A^{1/3}
\end{align}
where $p^+$ is the large light-cone momentum of the nucleons in the
nucleus, $A$ is the atomic number and $\Lambda$ is some transverse
momentum scale. In this case the expansion in powers of $\langle
T_{--} (x^-) \rangle$ translates into the expansion in the powers of
$\mu$, which can be systematically resummed (see e.g.
\cite{Grumiller:2008va}).

Alternatively one could argue that at strong coupling the
energy-momentum tensor is dominated not by valence quarks, but by the
strong gluon fields of the nucleus. One can argue, based on conformal
invariance, that the coordinate dependence of the energy-momentum
tensor of such a strong gluon field in ${\cal N} =4$ SYM theory is the
same as that of weakly coupled electromagnetic fields
\cite{Dominguez:2008vd,Chesler:2007sv,Gubser:2007nd}. Performing a
classical electrodynamics calculation of the energy-momentum tensor of
a point charge, averaging over all transverse coordinates and summing
over all nucleons yields (see Appendix \ref{A} for details)
\begin{align}\label{sea}
  \langle T_{--} (x^-) \rangle \, = \, \sqrt{\lambda} \, \Lambda^2 \,
  N_c^2 \, A^{1/3} \, \delta (x^{- \, 2})
\end{align}
with $\Lambda$ some transverse momentum scale. At weak coupling in
classical electrodynamics \\ $\langle T_{--} (x^-) \rangle \sim g^2
\sim \lambda$ with $\lambda$ the 't Hooft coupling
\begin{align}\label{thooft}
  \lambda \, = \, g^2 \, N_c.
\end{align}
Guided by the calculation of the heavy quark--antiquark potential at
strong coupling in \cite{Maldacena:1998im} we have replaced the
coupling constant $\lambda$ by $\sqrt{\lambda}$ in \eq{sea}. However,
the exact power of $\lambda$ in \eq{sea} does not alter the subsequent
discussion.

In case of the energy-momentum tensor in \eq{sea} one can construct an
expansion in the powers of $\Lambda^2$, which is again well-defined.

\FIGURE{\includegraphics[width=17cm]{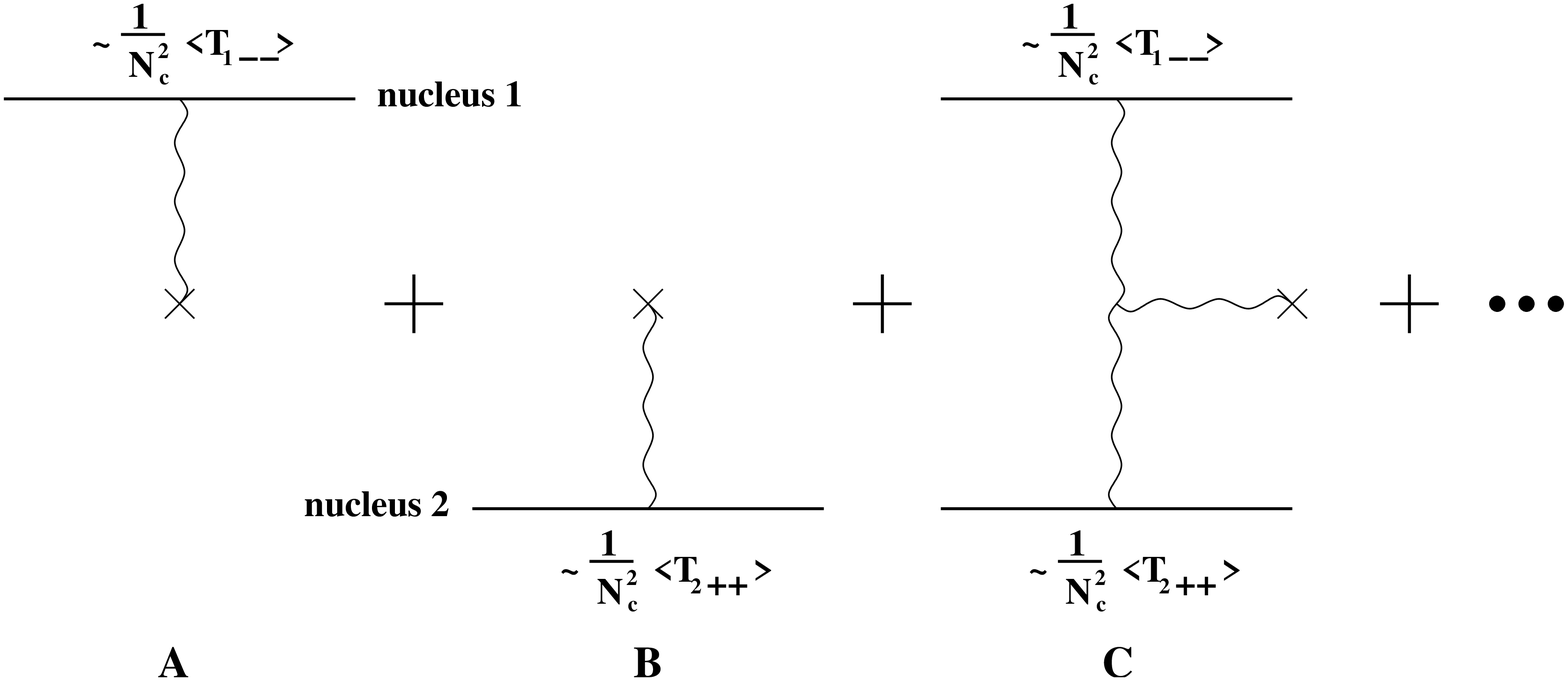}
  \caption{Diagrammatic representation of the metric in \protect\eq{2nuc1}. 
    Wavy lines are graviton propagators between the boundary of the
    AdS space and the bulk.  Graphs A and B correspond to the metrics
    of the first and the second nucleus correspondingly.  Diagram C is
    an example of the higher order graviton exchange corrections. We
    calculate the contribution of this diagram below in Sect.
    \protect\ref{gensol}.  }
  \label{pert}
}

Using the perturbative expansion in the powers of the energy-momentum
tensor, one can construct the metric dual to a heavy ion collision. At
the lowest non-trivial order we begin by writing the metric as
\begin{align}\label{2nuc1}
  ds^2 \, = \, \frac{L^2}{z^2} \, \bigg\{ -2 \, dx^+ \, dx^- + d
  x_\perp^2 + d z^2 + & \frac{2 \, \pi^2}{N_c^2} \, \langle T_{1 \,
    --} (x^-) \rangle \, z^4 \, d x^{- \, 2} + \frac{2 \,
    \pi^2}{N_c^2} \, \langle T_{2 \, ++} (x^+) \rangle \, z^4 \, d
  x^{+ \, 2} \notag \\ + & \, \text{higher order graviton exchanges}
  \bigg\}
\end{align}
where $\langle T_{1 \, --} (x^-) \rangle$ and $ \langle T_{2 \, ++}
(x^+) \rangle$ are the energy-momentum tensors of the colliding nuclei
1 and 2 as shown in \fig{spacetime}. The metric in \eq{2nuc1} is that
of two colliding shock waves in AdS$_5$. Higher order graviton
exchanges will modify the shock waves after the collision and generate
energy-momentum tensor of the matter produced by the collision in the
forward light cone. The metric of \eq{2nuc1} is our formulation of the
problem of heavy ion collisions in AdS. Similar metrics were
previously considered in modeling heavy ion collisions in AdS$_3$ in
\cite{Kajantie:2008rx} and in AdS$_5$ in
\cite{Nastase:2005rp,Grumiller:2008va}.

The metric of \eq{2nuc1} is illustrated in \fig{pert}. The first two
terms in \fig{pert} (diagrams A and B) correspond to one-graviton
exchanges which constitute the individual metrics of each of the
nuclei, as shown in \eq{1nuc}. Our goal below is to calculate the next
order correction to these terms, which is shown in the diagram C in
\fig{pert}. Higher order graviton exchanges would necessarily involve
both nuclei: as the metric (\ref{1nuc}) is an exact solution of
Einstein equations, all higher order graviton exchanges with a single
nucleus are zero. Indeed, solving Einstein equations order-by-order in
the graviton exchanges one could reconstruct any higher order term in
the series of \fig{pert}. In the calculations below we will restrict
ourselves to diagram C in \fig{pert}, which is the first correction to
the sum of the metrics of the two nuclei, and leave calculation of the
higher orders for future projects.


\section{General Perturbative Solution}
\label{gensol}

Here we will calculate the diagram C in \fig{pert} by solving Einstein
equations (\ref{ein2}) perturbatively. Define normalized light-cone
components of the nuclear energy-momentum tensors by
\begin{align}\label{t1}
  t_1 (x^-) \, \equiv \, \frac{2 \, \pi^2}{N_c^2} \, \langle T_{1 \,
    --} (x^-) \rangle
\end{align}
and
\begin{align}\label{t2}
  t_2 (x^+) \, \equiv \, \frac{2 \, \pi^2}{N_c^2} \, \langle T_{2 \,
    ++} (x^+) \rangle.
\end{align}
Using these definitions we rewrite the metric in \eq{2nuc1} as
\begin{align}\label{2nuc2}
  ds^2 \, = \, \frac{L^2}{z^2} \, \bigg\{ -2 \, dx^+ \, dx^- + d
  x_\perp^2 + d z^2 + t_1 (x^-) \, z^4 \, d x^{- \, 2} + t_2 (x^+) \,
  z^4 \, d x^{+ \, 2} + o (t_1 \, t_2) \bigg\}.
\end{align}
Notice that there is no higher order corrections containing only
powers of $t_1 (x^-)$ or of $t_2 (x^+)$: they are zero since the
single nucleus metric (\ref{1nuc}) and its analogue for nucleus 2 are
{\sl exact} solutions of Einstein equations.

We denote by $g_{\mu\nu}$ the metric in AdS$_5$ space dual to heavy
ion collisions that we are trying to construct. Then \eq{2nuc2} can be
written as
\begin{align}\label{metric}
  ds^2 \, = \, g_{\mu\nu} \, dx^\mu \, dx^\nu
\end{align}
with $\mu, \nu$ running from $0$ to $4$. The order-by-order
perturbative solution of Einstein equations is obtained by expanding
the metric around the empty AdS$_5$ space
\begin{align}\label{gpert}
  g_{\mu\nu} \, = \, g_{\mu\nu}^{(0)} + g_{\mu\nu}^{(1)} +
  g_{\mu\nu}^{(2)} + \ldots \, 
\end{align}
where the metric $g_{\mu\nu}^{(n)}$ corresponds to $n$ graviton
exchanges. For the energy-momentum tensor of \eq{mu} the series in
\eq{gpert} corresponds to expansion in the powers of $\mu$, while for
the energy-momentum tensor of \eq{sea} the series in \eq{gpert} is an
expansion in powers of $\Lambda^2$.

Here $g_{\mu\nu}^{(0)}$ is the metric of the empty AdS$_5$ space with
non-zero components
\begin{align}\label{metric0}
  g_{+-}^{(0)} = g_{-+}^{(0)} = - \frac{L^2}{z^2}, \ \ \ g_{ij}^{(0)}
  = \delta_{ij} \, \frac{L^2}{z^2}, \ \ i,j = 1, 2, \ \ \ \ \ 
  g_{zz}^{(0)} = \frac{L^2}{z^2}.
\end{align}
$g_{\mu\nu}^{(1)}$ is the first perturbation of the empty AdS$_5$
space due to the two nuclei
\begin{align}\label{metric1}
  g_{--}^{(1)} = t_1 (x^-) \, L^2 \, z^2, \ \ \ g_{++}^{(1)} = t_2
  (x^+) \, L^2 \, z^2
\end{align}
with all the other components zero. 

We want to find the next non-trivial correction $g_{\mu\nu}^{(2)}$. By
the choice of Fefferman-Graham coordinates one has $g_{z \mu}= g_{\mu
  z} = 0$ exactly for $\mu \neq z$ and $g_{zz} = L^2 / z^2$. Hence the
non-trivial components of $g_{\mu\nu}^{(2)}$ are those for $\mu, \nu =
0, \ldots, 3$. Due to translational and rotational invariance of the
nuclei in the transverse direction $g_{ij}^{(2)} \sim \delta_{ij}$. We
thus parametrize the unknown components of $g_{\mu\nu}^{(2)}$ as
\begin{align}\label{metric2}
  & g_{--}^{(2)} = \frac{L^2}{z^2} \, f (x^+, x^-, z), \ \ \ 
  g_{++}^{(2)} = \frac{L^2}{z^2} \, {\tilde f} (x^+, x^-, z), \notag \\
  & g_{+-}^{(2)} = - \frac{1}{2} \, \frac{L^2}{z^2} \, g (x^+, x^-,
  z), \ \ \ g_{ij}^{(2)} = \frac{L^2}{z^2} \, h (x^+, x^-, z) \,
  \delta_{ij}
\end{align}
with $f$, $\tilde f$, $g$ and $h$ some unknown functions. Imposing
causality we require that functions $f$, $\tilde f$, $g$ and $h$ are
zero before the collision, i.e., that before the collision the metric
is given only by the empty AdS space and by the contributions of the
two nuclei \eq{metric1}.  Also, according to general properties of
$g_{\mu\nu}$ outlined in Sect. \ref{general} (see
\cite{deHaro:2000xn}), we demand that $f$, $\tilde f$, $g$ and $h$ go
to zero as $z^4$ when $z \rightarrow 0$.

Using Eqs. (\ref{metric0}), (\ref{metric1}), and (\ref{metric2}) in
\eq{gpert}, plugging the latter into Einstein equations (\ref{ein2})
and keeping only the terms up to and including the order
$g_{\mu\nu}^{(2)}$ we obtain the following set of equations for $f$,
$\tilde f$, $g$ and $h$ labeled by the Einstein equations components:
 
\begin{subequations}\label{ein}
\begin{align}
  (--) \hspace*{1cm} & \frac{3}{2 \, z} \, f_{z} - \frac{1}{2} \, f_{z
    \, z} -
  h_{x^- \, x^-} = 0, \label{--} \\
  (++) \hspace*{1cm} & \frac{3}{2 \, z} \, {\tilde f}_{z} -
  \frac{1}{2} \, {\tilde f}_{z \, z} -
  h_{x^+ \, x^+} = 0, \label{++} \\
  (+-) \hspace*{1cm} & - \frac{5}{4 \, z} \, g_z - \frac{1}{z} \, h_z
  + \frac{1}{4} \, g_{z \, z} - \frac{1}{2} \, g_{x^+ \, x^-} - h_{x^+
    \, x^-} - \frac{1}{2} \, f_{x^+ \, x^+} - \frac{1}{2} \, {\tilde
    f}_{x^- \, x^-} \notag \\ & = 4 \, z^6 \, t_1 (x^-) \,
  t_2 (x^+)  - \frac{1}{4} \, z^8 \, t'_1 (x^-) \, t'_2 (x^+), \label{+-} \\
  (\perp \perp) \hspace*{1cm} & g_z + 5 \, h_z - z \, h_{z \, z} + 2
  \, z \, h_{x^+ \, x^-} = 8 \,
  z^7 \, t_1 (x^-) \, t_2 (x^+), \label{pp} \\
  (zz) \hspace*{1cm} & g_z + 2 \, h_z - z \, g_{z \, z} - 2 \, z \,
  h_{z \, z} = - 32 \,
  z^7 \, t_1 (x^-) \, t_2 (x^+), \label{zz} \\
  (-z) \hspace*{1cm} & - \frac{1}{2} \, f_{x^+ \, z} - \frac{1}{4} \,
  g_{x^- \, z} -
  h_{x^- \, z} = - z^7 \, t'_1 (x^-) \, t_2 (x^+),  \label{-z} \\
  (+z) \hspace*{1cm} & - \frac{1}{2} \, {\tilde f}_{x^- \, z} -
  \frac{1}{4} \, g_{x^+ \, z} - h_{x^+ \, z} = - z^7 \, t_1 (x^-) \,
  t'_2 (x^+).
  \label{+z}
\end{align}
\end{subequations}
The subscripts $z$, $x^+$ and $x^-$ indicate partial derivatives with
respect to these variables.

To solve Eqs. (\ref{ein}) begin by solving \eq{pp} for $g_z$ and
substituting the result into \eq{zz}. This gives
\begin{align}\label{heq}
  - 3 \, h_z + 3 \, z \, h_{z \, z} - z^2 \, h_{z \, z \, z} + 2 \,
  z^2 \, h_{x^+ \, x^- \, z} = 16 \, z^7 \, t_1 (x^-) \, t_2 (x^+).
\end{align}
We look for the solution of \eq{heq} as a series in powers of $z^2$.
Note that $h (x^+, x^-, z)$ goes to zero proportionally to $z^4$ as $z
\rightarrow 0$: therefore the series starts at the order $z^4$ and
reads
\begin{align}\label{hser}
  h (x^+, x^-, z) \, = \, z^4 \, \sum\limits_{n=0}^\infty h_n (x^+,
  x^-) \, z^{2n}.
\end{align}
Substituting \eq{hser} into \eq{heq} and solving it order-by-order in
$z$ we can express all the coefficients in the series in terms of the
first coefficient $h_0 (x^+, x^-)$ and in terms of $t_1 (x^-)$ and
$t_2 (x^+)$ obtaining
\begin{align}\label{hsol_long}
  h (x^+, x^-, z) \, = \, 4 \, z^2 \, \frac{I_2 (z \sqrt{2 \,
      \partial_+ \, \partial_-})}{\partial_+ \, \partial_-} \ h_0
  (x^+, x^-) - 32 \, z^2 \, \bigg[ I_2 (z \sqrt{2 \, \partial_+ \,
    \partial_-}) \notag \\ - \frac{1}{4} \, z^2 \, \partial_+ \,
  \partial_- - \frac{1}{24} \, z^4 \, (\partial_+ \, \partial_-)^2
  \bigg] \, \frac{1}{(\partial_+ \, \partial_-)^3} \ t_1 (x^-) \, t_2
  (x^+).
\end{align}
(Inverse derivatives in \eq{hsol_long} are canceled by the positive
powers of derivatives in the numerators of the appropriate terms.)

As shown in Appendix \ref{B}, plugging \eq{hsol_long} into \eq{pp} one
can find $g (x^+, x^-, z)$, and, using Eqs. (\ref{-z}) and (\ref{+z}),
one can find $f (x^+, x^-, z)$ and ${\tilde f} (x^+, x^-, z)$. One can
argue (see Appendix \ref{B}) that for the solution of Einstein
equations to satisfy initial ($h=g=f={\tilde f}=0$ before the
collision) and boundary ($h (x^+, x^-, z=0)=0$) conditions the
following relation needs to be satisfied:
\begin{align}\label{h0eq}
  (\partial_+ \, \partial_-)^2 \, h_{0} (x^+, x^-) \, = \, 8 \, t_1
  (x^-) \, t_2 (x^+).
\end{align}
As can be seen from \eq{hsol_long} the infinite series (\ref{hser})
for $h (x^+, x^-, z)$ will then terminate at the order $z^6$. As shown
in Appendix \ref{B}, the solutions for $f (x^+, x^-, z)$, ${\tilde f}
(x^+, x^-, z)$ and $g (x^+, x^-, z)$ will also reduce to finite-order
polynomials in $z^2$.

The only other non-vanishing coefficient in the series for $h (x^+,
x^-, z)$ in \eq{hser} is $h_1 (x^+, x^-)$ which is related to $h_{0}
(x^+, x^-)$ via
\begin{align}\label{h1eq}
  h_1 (x^+, x^-) \, = \, \frac{1}{6} \, \partial_+ \, \partial_- \,
  h_{0} (x^+, x^-).
\end{align}
Using \eq{h0eq} we obtain
\begin{align}\label{h1eq2}
  \partial_+ \, \partial_- \, h_1 (x^+, x^-) \, = \, \frac{4}{3} \,
  t_1 (x^-) \, t_2 (x^+).
\end{align}

Eqs. (\ref{h0eq}) and (\ref{h1eq2}) allow us to determine the
functions $h_0$ and $h_1$. Imposing causality by requiring that at
time $-\infty$, i.e. long before the collision, the shock waves are
unmodified we write
\begin{align}\label{h0int}
  h_0 (x^+, x^-) \, = \, 8 \, \int\limits_{-\infty}^{x^-} dx'^- \,
  \int\limits_{-\infty}^{x'^-} dx''^- \, \int\limits_{-\infty}^{x^+}
  dx'^+ \, \int\limits_{-\infty}^{x'^+} dx''^+ \, t_1 (x''^-) \, t_2
  (x''^+)
\end{align}
and
\begin{align}\label{h1int}
  h_1 (x^+, x^-) \, = \, \frac{4}{3} \, \int\limits_{-\infty}^{x^-}
  dx'^- \, \int\limits_{-\infty}^{x^+} dx'^+ \, t_1 (x'^-) \, t_2
  (x'^+).
\end{align}
In terms of $h_0$ and $h_1$ from Eqs. (\ref{h0int}) and (\ref{h1int})
we write our solution for $h (x^+, x^-, z)$ as
\begin{align}\label{hsol}
  h (x^+, x^-, z) \, = \, h_0 (x^+, x^-) \, z^4 + h_1 (x^+, x^-) \,
  z^6.
\end{align}
Plugging the solution (\ref{hsol}) into \eq{pp} we solve for $g (x^+,
x^-, z)$ to obtain
\begin{align}\label{gsol}
  g (x^+, x^-, z) \, = \, - 2 \, h_0 (x^+, x^-) \, z^4 - 2 \, h_1
  (x^+, x^-) \, z^6 + \frac{2}{3} \, t_1 (x^-) \, t_2 (x^+) \, z^8.
\end{align}
Substituting the solutions for $h$ and $g$ from Eqs. (\ref{hsol}) and
(\ref{gsol}) into \eq{-z} we solve for $f (x^+, x^-, z)$ to find
\begin{align}\label{fsol}
  f (x^+, x^-, z) \, = \, - \lambda_1 (x^+, x^-) \, z^4 - \frac{1}{6}
  \, \partial_-^2 h_0 (x^+, x^-) \, z^6 - \frac{1}{16} \, \partial_-^2
  h_1 (x^+, x^-) \, z^8
\end{align}
with $\lambda_1 (x^+, x^-)$ given by
\begin{align}\label{lam}
  \lambda_1 (x^+, x^-) \, = \, \int\limits_{-\infty}^{x^+} d x'^+ \,
  \partial_- h_0 (x'^+, x^-).
\end{align}
Similarly Eq. (\ref{+z}) yields
\begin{align}\label{fdsol}
  {\tilde f} (x^+, x^-, z) \, = \, - \lambda_2 (x^+, x^-) \, z^4 - \frac{1}{6}
  \, \partial_+^2 h_0 (x^+, x^-) \, z^6 - \frac{1}{16} \, \partial_+^2
  h_1 (x^+, x^-) \, z^8
\end{align}
with 
\begin{align}\label{lam2}
  \lambda_2 (x^+, x^-) \, = \, \int\limits_{-\infty}^{x^-} d x'^- \,
  \partial_+ h_0 (x^+, x'^-).
\end{align}

Eqs. (\ref{hsol}), (\ref{gsol}), (\ref{fsol}) and (\ref{fdsol})
provide us with the solution of Eq. (\ref{ein}) with the causal
initial condition requiring all these functions to go to zero at
infinitely early times.

Using \eq{holo} one can obtain the contribution of $g_{\mu\nu}^{(2)}$
to the expectation value of the energy-momentum tensor at the boundary
of the AdS space from Eqs.  (\ref{hsol}), (\ref{gsol}), (\ref{fsol})
and (\ref{fdsol}) and \eq{metric2}:
\begin{align}\label{Tsol}
  \langle T_{- \, -} \rangle \, = \, - \frac{N_c^2}{2 \, \pi^2} \,
  \lambda_1 (x^+, x^-), \ \ \ \langle T_{+ \, +} \rangle \, = \, -
  \frac{N_c^2}{2 \, \pi^2} \, \lambda_2 (x^-, x^+), \notag \\ \langle
  T_{+ \, -} \rangle \, = \, \frac{N_c^2}{2 \, \pi^2} \, h_0 (x^-,
  x^+), \ \ \ \langle T_{i \, j} \rangle \, = \, \frac{N_c^2}{2 \,
    \pi^2} \, \delta_{i \, j} \, h_0 (x^-, x^+).
\end{align}
Given $t_1 (x^-)$ and $t_2 (x^+)$, one can use Eqs.  (\ref{h0int}),
(\ref{lam}) and (\ref{lam2}) to find $h_0$, $\lambda_1$ and
$\lambda_2$, and then use \eq{Tsol} to construct the energy-momentum
tensor of the gauge theory.


\section{Physical Shock Waves: Nuclear Stopping}
\label{real_shock}

To understand our solution given by Eqs. (\ref{hsol}), (\ref{gsol}),
(\ref{fsol}) and (\ref{fdsol}) let us consider a specific example of
shock waves with the boundary energy-momentum tensor given by
\eq{Tmu}. To be able to better understand physical properties of the
solution, let us ``smear'' the shock waves over some finite
longitudinal distance. If one imagines shock waves representing a
large nucleus, such nucleus moving in the $x^+$-direction in a boosted
ultrarelativistic frame would have a longitudinal extent
\begin{align}\label{a}
  a \, \propto \, R \, \frac{\Lambda}{p^+} \, \propto \, \frac{A^{1/3}}{p^+}
\end{align}
with $R$ the nuclear radius, $\Lambda$ the typical transverse momentum
scale in the problem ($R \propto A^{1/3}/\Lambda$ with $A$ the atomic
number), and $p^+$ the large longitudinal momentum of the nucleus.
(Here $\Lambda/p^+$ is the boost factor.)

Assuming that the nucleus has equal thickness $a$ at all impact
parameters, we replace the delta-function in \eq{Tmu} with two
theta-functions to write
\begin{align}\label{t1s}
  t_1 (x^-) \, = \, 2 \, \pi^2 \, \frac{\mu}{a} \, \theta (x^-) \,
  \theta (a - x^-)
\end{align}
for the first nucleus and
\begin{align}\label{t2s}
  t_2 (x^+) \, = \, 2 \, \pi^2 \, \frac{\mu}{a} \, \theta (x^+) \,
  \theta (a - x^+)
\end{align}
for the second one. For simplicity we assumed that the nuclei are
identical and are scattering with equal momenta $p_1^+ = p_2^-$, such
that $\mu = \mu_1 = \mu_2$ and $a = a_1 = a_2$. 

Plugging Eqs. (\ref{t1s}) and (\ref{t2s}) into \eq{h0int} we
immediately obtain
\begin{align}
  h_0 (x^+, x^-) \, = \, 8 \, \frac{\mu^2}{a^2} \, (2 \, \pi^2)^2 \, &
  \bigg[ \theta (x^-) \, \theta (a - x^-) \, \frac{x^{- \, 2}}{2} +
  \theta (x^- - a) \, a \, \left( x^- - \frac{a}{2} \right) \bigg]
  \notag \\ & \bigg[ \theta (x^+) \, \theta (a - x^+) \, \frac{x^{+ \,
      2}}{2} + \theta (x^+ - a) \, a \, \left( x^+ - \frac{a}{2}
  \right) \bigg]. \label{h0s}
\end{align}
\eq{lam} then gives
\begin{align}\label{lam1s}
  \lambda_1 (x^+, x^-) \, = \, 8 \, \frac{\mu^2}{a^2} \, (2 \, \pi^2)^2
  \, & \bigg[ \theta (x^-) \, \theta (a - x^-) \, x^- + \theta (x^- -
  a) \, a \bigg] \, \notag \\ & \bigg[ \theta (x^+) \, \theta (a -
  x^+) \, \frac{x^{+ \, 3}}{6} + \theta (x^+ - a) \, a \, \left(
    \frac{a^2}{6} + \frac{x^{+\, 2}}{2} - \frac{a \, x^+}{2} \right)
  \bigg],
\end{align}
while \eq{lam2} due to the fact that nuclei are identical leads to
\begin{align}\label{lam2s}
  \lambda_2 (x^+, x^-) \, = \, \lambda_1 (x^-, x^+).
\end{align}
Eqs. (\ref{h0s}), (\ref{lam1s}) and (\ref{lam2s}), along with
\eq{Tsol}, give us the order $\mu^2$ energy-momentum tensor.  Let us
study its properties. First of all, away from the light cone for $x^+,
x^- \gg a$ (or in the limit of infinitely thin nuclei, which can be
recovered by taking $a \rightarrow 0$) one has
\begin{align}\label{far1}
  h_0 (x^+, x^-) \bigg|_{x^+, x^- \gg a} \, \approx \, 8 \, (2 \,
  \pi^2)^2 \, \mu^2 \, x^+ \, x^-, \ \ \ \lambda_1 (x^+, x^-)
  \bigg|_{x^+, x^- \gg a} \, \approx \, 8 \, (2 \, \pi^2)^2 \, \mu^2
  \, \frac{x^{+\, 2}}{2}, \notag \\ \lambda_2 (x^+, x^-) \bigg|_{x^+,
    x^- \gg a} \, \approx \, 8 \, (2 \, \pi^2)^2 \, \mu^2 \,
  \frac{x^{-\, 2}}{2}.
\end{align}
Substituting \eq{far1} into \eq{Tsol} one gets for the forward
light-cone far away from the nuclei
\begin{align}\label{Tfar}
  \langle T_{- \, -} \rangle \, = \, - 8 \, \pi^2 \, N_c^2 \, \mu^2 \,
  x^{+\, 2}, \ \ \ \langle T_{+ \, +} \rangle \, = \, - 8 \, \pi^2 \,
  N_c^2 \, \mu^2 \, x^{- \, 2}, \notag \\ \langle T_{+ \, -} \rangle
  \, = \, 8 \, \pi^2 \, N_c^2 \, \mu^2 \, \tau^2, \ \ \ \langle T_{i
    \, j} \rangle \, = \, 8 \, \pi^2 \, \delta_{i \, j} \, N_c^2 \,
  \mu^2 \, \tau^2.
\end{align}
This is the same result as obtained in \cite{Grumiller:2008va}. The
energy-momentum tensor in \eq{Tfar} is rapidity-independent, as its
components contain no rapidity dependence apart from the trivial
factors needed for Lorentz-properties of the tensor.

For two colliding nuclei of infinite transverse extent (the Bjorken
case \cite{Bjorken:1982qr}) the most general parameterization of the
rapidity-independent energy-momentum tensor is \cite{Kovchegov:2005ss}
\begin{align}\label{tmngen}
& T_{--} \, = \, [\epsilon (\tau) + p_3 (\tau)] \, \left( \frac{x^+}{\tau}
\right)^2, \nonumber ~\\ 
& T_{++} \, = \, [\epsilon (\tau) + p_3 (\tau)] \, \left( 
\frac{x^-}{\tau} \right)^2, \nonumber ~\\ 
& T_{+-} \, = \,  [\epsilon (\tau) - p_3 (\tau)] \, \frac{1}{2}, \nonumber ~\\ 
& T_{ij} \, = \, \delta_{ij} \, p (\tau),
\end{align}
where $\epsilon (\tau)$, $p (\tau)$ and $p_3 (\tau)$ are the energy
density, transverse pressure and longitudinal pressure components of
the energy-momentum tensor at mid-rapidity ($x^3=0$). At $x^3=0$ the
tensor (\ref{tmngen}) looks like
\begin{align}
   T^{\mu\nu} =  
 \left( 
\begin{matrix}
\epsilon (\tau) & 0 & 0 & 0 \\
  0 & p (\tau) & 0 & 0 \\
  0 & 0 & p (\tau) & 0  \\
  0 & 0 & 0  & p_3 (\tau) \\ 
\end{matrix}
\right)\, .
\end{align}

One can easily show that conservation of energy and momentum condition
\begin{align}
  \partial_\mu T^{\mu\nu} \, = \, 0
\end{align}
applied to the tensor (\ref{tmngen}) gives
\begin{align}\label{bjhyd}
  \frac{d \epsilon}{d \tau} \, = \, - \frac{\epsilon + p_3}{\tau}.
\end{align}
The condition $\partial_\mu T^{\mu\nu} = 0$ follows from Einstein
equations if one uses gauge-gravity duality to obtain the energy
momentum tensor. For conformal field theories the energy-momentum
tensor is traceless
\begin{align}
  T_\mu^\mu \, =\, 0,
\end{align}
which implies
\begin{align}\label{trless}
  \epsilon \, = \, 2 \, p + p_3.
\end{align}
Eqs. (\ref{bjhyd}) and (\ref{trless}) relate $\epsilon (\tau)$, $p
(\tau)$ and $p_3 (\tau)$ to each other, such that knowing one of these
functions is sufficient to reconstruct the others.

Comparing \eq{Tfar} with \eq{tmngen} we read off the energy density at
mid-rapidity
\begin{align}\label{ereal}
  \epsilon (\tau) \, = \, 4 \, \pi^2 \, N_c^2 \, \mu^2 \, \tau^2.
\end{align}
Such energy density at mid-rapidity is problematic. It grows with the
proper time $\tau$. One can show that the requirement that the energy
density of produced matter is positive-definite in any frame in
particular demands that \cite{Janik:2005zt}
\begin{align}\label{pos}
  \epsilon' (\tau) \, \le \, 0.
\end{align}
\eq{ereal} obviously violates the condition (\ref{pos}): this means
our solution gives {\sl negative energy density} in some frames. Such
result is clearly unphysical. 

It is important to understand the origin of this negativity of the
energy density. First we note that, as one can easily see, the energy
density becomes negative in the frames with the time direction being
close to the light cones (the shock waves). To investigate the region
around the shock waves further, let us concentrate on the shock wave
corresponding to the nucleus 1 after the collision. Let us study what
happens to, say, the middle of the nucleus, which is located at $x^- =
a/2$, after the collision. The important component of the
energy-momentum tensor is $\langle T_{- \, -} \rangle$, since it
contains the (large) momentum component of the nucleus along its light
cone. Using \eq{Tsol} along with \eq{lam1s} at $x^- = a/2$ yields for
$x^+ \gg a$ (after the collision)
\begin{align}\label{stop}
  \langle T_{- \, -} (x^+ \gg a, x^- = a/2) \rangle \, = \, N_c^2 \,
  \frac{\mu}{a} - 4 \, \pi^2 \, N_c^2 \, \mu^2 \, x^{+\, 2},
\end{align}
where the first term on the right is due to the original shock wave
obtained by using Eqs. (\ref{t1s}) and (\ref{t1}) at $x^- = a/2$.

\eq{stop} shows that $\langle T_{- \, -} \rangle$ of a nucleus becomes
{\sl negative} at light-cone times 
\begin{align}\label{stoptime}
  x^+ \, \sim \, \frac{1}{\sqrt{\mu \, a}}. 
\end{align}
Indeed zero $\langle T_{- \, -} \rangle$ would mean a complete {\sl
  stopping} of the shock wave and the corresponding nucleus. We
therefore conclude that negativity of energy density (\ref{ereal}) in
fact is a signal of complete stopping of the colliding nuclei after
the collision!

Indeed at times $x^+ \, \sim \, 1/\sqrt{\mu \, a}$ higher order
corrections to the metric due to higher graviton exchanges would
become important preventing $\langle T_{- \, -} \rangle$ from going
negative. Nevertheless, \eq{stop} demonstrates that at rather short
times $x^+ \, \sim \, 1/\sqrt{\mu \, a}$ the nucleus looses the amount
of energy comparable to its initial incoming energy, and thus is
likely to stop.

One should also point out that \eq{stop} gives $\langle T_{- \, -}
\rangle$ of the center of the nucleus ($x^- = a/2$): other slices of
the nucleus located at different $x^-$ would also stop, but at
slightly different times $x^+$. All stopping would happen at the same
parametric time given by \eq{stoptime}.

To better understand the stopping time we use Eqs. (\ref{mu}) and
(\ref{a}) to re-write \eq{stoptime} as
\begin{align}\label{stoptime2}
  x^+ \, \sim \, \frac{1}{\Lambda \, A^{1/3}}. 
\end{align}
The stopping time appears to be energy-independent! It is given by the
inverse of the typical transverse momentum scale $\Lambda$ in the
problem. It also decreases with the increasing size of the nucleus
$A$.

Let us pause to interpret the main result of this Section. It appears
that two colliding ultrarelativistic shock waves would come to a
complete stop shortly after the collision. One can understand this in
terms of creation of a black hole: both shock waves carry large
energy, which functions as a mass. There is a horizon radius in AdS
space corresponding to that mass/energy. For nuclei of infinite
transverse extent under consideration the shock waves always come
closer to each other than the horizon radius corresponding to the
energy they carry.  A black hole is then formed and the shock waves
stop completely within the black hole's horizon radius. The picture is
similar to black hole production in collisions at transplanckian
energies, which has been recently discussed in the literature
\cite{Giddings:2001bu,Eardley:2002re,Kancheli:2002nw}.

One could picture a collision of two nuclei of finite transverse
extent. If the impact parameter of such a collision is larger than the
horizon radius, no black hole will be formed and the nuclei will not
stop. However, in such case there will probably be no thermal matter
produced in the boundary theory either.

It is interesting to note that the stopping time (\ref{stoptime2}) is
independent of energy: indeed, on one hand if one increases the
momentum of the shock wave it is harder to stop it. On the other hand,
increasing the energy of the shock waves tends to reduce the radius of
the horizon, trying to make the shock waves stop faster. We interpret
the result of \eq{stoptime2} as the cancellation of the two effects,
leading to energy-independence of the stopping time.

If the nuclei stop completely in the collision, the strong
interactions between them are almost certain to thermalize the system.
Indeed if the interactions were strong enough to stop the nuclei, they
should be strong enough to thermalize the resulting medium. The
dynamics of such a rotationally-invariant thermal medium was
originally described hydrodynamically by Landau in
\cite{Landau:1953gs} and is commonly referred to as Landau
hydrodynamics. Hence our conclusion is that modeling a collision of
two nuclei by two physical colliding shock waves in AdS necessarily
leads to complete nuclear stopping, and probably to thermalization of
the system and the subsequent dynamics describable by Landau
hydrodynamics.

The fact that nuclei do not stop instantaneously, but require certain
(short) time (\ref{stoptime2}) to stop avoids the standard
counter-argument \cite{Anishetty:1980zp} against Landau hydrodynamics
\cite{Landau:1953gs}, which suggests that it would violate the
uncertainty principle if the stopping was instantaneous.  Hence the
picture is intrinsically consistent.

One may wonder whether our result of \eq{stop} is specific to the
shape of the shock waves we have considered in Eqs. (\ref{t1s}) and
(\ref{t2s}). In fact our conclusion of complete stopping is valid in
general: as one can see from \eq{h0int}, any non-negative
energy-momentum tensor of the shock wave, which is positive in a
localized region of $x^-$ ($x^+$) axis, would give $h_0 \sim x^+ \,
x^-$ at late times. Eqs. (\ref{Tsol}) and (\ref{tmngen}) would then
give
\begin{align}\label{tau2}
  \epsilon (\tau) \, \sim \, p (\tau) \, \sim \, - p_3 (\tau) \, \sim \,
  \tau^2,
\end{align}
just like in our case considered above. Hence the energy density of
such system would never be non-negative in all frames, signaling the
stopping of shock waves. Finally, \eq{lam} would give $\lambda_1 \sim
x^{+ \, 2}$, such that the correction to the energy-momentum tensor on
the light cone would again be
\begin{align}
  \langle T_{- \, -} \rangle \, \sim \, - x^{+ \, 2}
\end{align}
indicating that at some large enough light-cone time $x^+$ the
nucleus would run out of its momentum. This proves the shock waves
stopping independent of the shape of the shock waves profiles.

Let us close this Section by pointing out that the result in
\eq{ereal} can be easily obtained (at the parametric level) for
infinitely thin nuclei by noticing that the diagram in \fig{pert}C,
which contributes to the metric giving the energy density in
\eq{ereal}, is of the order of $\mu^2$. Hence the contribution to
$\epsilon$ at this order should be of the order of $\mu^2$. However,
energy density has dimension of mass to the fourth power, while
$\mu^2$ has dimensions of mass to the sixth power. To make the
dimensions right we use the only other dimensionful quantity in the
boundary gauge theory in the forward light cone: the proper time
$\tau$. This gives $\epsilon \sim \mu^2 \tau^2$, in agreement with
\eq{ereal}. We noted above that expansion in graviton exchanges in the
bulk is equivalent to expansion in the powers of $\mu$ for the
energy-momentum tensor of the shock waves in \eq{Tmu}. The only way to
make a dimensionless expansion parameter from $\mu$ in the boundary
theory is to multiply it by $\tau^3$. Now we see that for energy
density the expansion parameter is in fact $\mu \, \tau^3$, which has
been noticed in \cite{Grumiller:2008va} before. However, now we
understand that each power of this expansion parameter corresponds to
a graviton exchange between the boundary and the bulk.


\section{Unphysical Shock Waves: Energy Density of the Produced Medium}
\label{our_shock}

In the above Section we came to the conclusion that at very strong
coupling colliding nuclei completely stop in a collision (for central
collisions), forming a medium described by Landau hydrodynamics.
However, due to asymptotic freedom of QCD, we know that small-coupling
effects play an important role in heavy ion collisions and in high
energy collisions in general. The success of Color Glass Condensate
\cite{Iancu:2003xm,Weigert:2005us,Jalilian-Marian:2005jf} based models
in describing RHIC data (see e.g. \cite{McLerran:2008uj} and
references therein) suggests that weakly coupled effects are present
in the actual heavy ion collisions at RHIC, at least at very early
times during and after the collision. While a comprehensive
description of both weakly-coupled initial dynamics and
strongly-coupled dynamics of the produced medium is not feasible at
this point, here we will suggest a model capturing some of the feature
of the weakly coupled collisions.

We begin by noting that, in the weak coupling limit, the colliding
nuclei do not stop, as we observed in the previous Section for the
strong coupling case. Instead the valence quarks and other large
Bjorken-$x$ (hard) partons are usually assumed to go through each
other without deflection, shedding off the softer (small-$x$) virtual
partons, which are left behind and quickly go on mass shell, i.e.,
become real
\cite{Kovner:1995ts,Kovner:1995ja,Kovchegov:1997ke,Gyulassy:1997vt,Krasnitz:2003nv,Krasnitz:2003jw,Krasnitz:1999wc,Krasnitz:2002mn,Kovchegov:2000hz,Baier:2000sb}.
The medium made out of these small-$x$ partons after the collision has
a non-negative energy density in any frame
\cite{Lappi:2003bi,Fukushima:2007ja,Krasnitz:2002mn,Krasnitz:2003nv,Krasnitz:2003jw}.
We will then proceed by requiring that the energy density of the
produced strongly coupled medium should also be non-negative.

We want to model the heavy ion collisions by colliding two shock
waves. The conclusion of the previous section was that any
localized non-negative $\langle T_{1 \, - \, -} \rangle$ of a shock wave,
such that
\begin{align}
  \int\limits_{-\infty}^\infty \, d x^- \langle T_{1 \, - \, -} (x^-)
  \rangle \, > \, 0
\end{align}
(with an analogous condition imposed on the $\langle T_{2 \, + \, +}
(x^+) \rangle$ component of the energy momentum tensor of the other
shock wave) leads to the energy density scaling of \eq{tau2}. This
violates the condition (\ref{pos}) derived in \cite{Janik:2005zt} and
results in the negative energy density of the produced matter in some
frames.  The only way around such an unphysical conclusion appears to
be to require that
\begin{align}\label{zero}
  \int\limits_{-\infty}^\infty \, d x^- \langle T_{1 \, - \, -} (x^-)
  \rangle \, = \, 0, \ \ \ \int\limits_{-\infty}^\infty \, d x^+
  \langle T_{2 \, + \, +} (x^+) \rangle \, = \, 0.
\end{align}
Indeed the conditions (\ref{zero}) can only be satisfied in a physical
world if there is no shock waves, in which case their energy would be
zero. Such a trivial scenario is not what we have in mind. 

Instead, we propose using unphysical not positive-definite quantities
for $\langle T_{1 \, - \, -} (x^-) \rangle$ and \\ $\langle T_{2 \, +
  \, +} (x^+) \rangle$, which integrate out to zero satisfying
\eq{zero}.  Indeed such objects would be completely unphysical, as
they would contain regions of negative energy density. They can not be
obtained from an underlying string theory either. However, we intend
to use them in gauge-gravity duality only. On both sides of the
gauge-gravity duality our non-positive energy momentum tensors should
be regarded as external sources to the theory.  The conclusion we have
reached is that to have non-negative energy density in the forward
light cone one needs unphysical negative energy shock waves on the
light cone itself.

Indeed our proposal of zero-energy shock waves may not be a unique way
of modeling weak coupling effects in heavy ion collisions in the
AdS/CFT framework. One may also try to construct a metric with the
CGC-inspired energy-momentum tensor for the gauge theory at early
proper time and evolve it in time using Einstein equations. However,
constructing a metric which is a valid initial condition for Einstein
equations at early times and accounts for perturbative features of the
collision appears to be difficult. If one insists on modeling the
heavy ion collisions by two colliding shock waves, our zero-energy
shock wave proposal is the only way to mimic the weak coupling effects
at initial stages of the collision.

Inspired by Eqs. (\ref{Tmu}) and (\ref{sea}), which contain two
factors of transverse momenta times some function of longitudinal
coordinates and momenta, we suggest describing the energy-momentum
tensors of the colliding nuclei by
\begin{align}\label{Tzero}
  \langle T_{1 \, - \, -} (x^-) \rangle \, = \, \frac{N_c^2}{2 \,
    \pi^2} \,
  \Lambda_1^2 \, \delta' (x^-) \notag \\
  \langle T_{2 \, + \, +} (x^+) \rangle \, = \, \frac{N_c^2}{2 \,
    \pi^2} \, \Lambda_2^2 \, \delta' (x^+)
\end{align}
corresponding to
\begin{align}\label{tzero}
  t_1 (x^-) \, = \, \Lambda_1^2 \, \delta' (x^-) \notag \\
  t_2 (x^+) \, = \, \Lambda_2^2 \, \delta' (x^+)
\end{align}
in the shock waves metric in \eq{2nuc2}. $\delta'(x)$ denotes the
derivative of a delta-function. Clearly the energy-momentum tensors in
\eq{Tzero} satisfy \eq{zero}. What we loose in this description is the
relation between the transverse momentum scales $\Lambda_1^2$ and
$\Lambda_2^2$ describing the two nuclei in \eq{Tzero} and the actual
physical parameters describing the real nuclei, since our
energy-momentum tensors in \eq{Tzero} are not physical and we can not
relate them to the energy-momentum tensors of the two nuclei.

Before we perform any calculations we can already guess the answer
using the dimensional analysis outlined in Sect. \ref{real_shock}.
This time each vertex in \fig{pert}C brings in a factor of
$\Lambda_1^2$ and $\Lambda_2^2$, such that the diagram is proportional
to $\Lambda_1^2 \, \Lambda_2^2$. Hence the resulting energy density of
the boundary theory is proportional to $\epsilon \sim \Lambda_1^2 \,
\Lambda_2^2$. Since the dimensions of $\epsilon$ and $\Lambda_1^2 \,
\Lambda_2^2$ match, no powers of $\tau$ are needed this time. Hence we
conclude that the energy density of the matter produced by the two
shock waves (\ref{tzero}) at the lowest order in graviton exchanges is
$\epsilon \sim \Lambda_1^2 \, \Lambda_2^2$, i.e. a {\sl constant} of
time, as was suggested in \cite{Kovchegov:2007pq}. $\epsilon \sim
\Lambda_1^2 \, \Lambda_2^2$ immediately satisfies the condition
(\ref{pos}) derived in \cite{Janik:2005zt}: hence the energy density
is non-negative in any reference frame. Finally, now the graviton
exchanges between the boundary and the bulk should correspond to
powers of $\Lambda_1^2 \, \tau^2$ or $\Lambda_2^2 \, \tau^2$ in the
gauge theory. Thus the early-time expansion for the energy density
should contain powers of $\Lambda_1^2 \, \tau^2$ and $\Lambda_2^2 \,
\tau^2$. Therefore, while we do not know how to relate $\Lambda_1^2$
and $\Lambda_2^2$ to the physical observables, we still can
systematically construct the dual geometry to the collision by
expanding the metric in the powers of $\Lambda_1^2 \, \tau^2$ and
$\Lambda_2^2 \, \tau^2$, and hopefully would be able to arrive at the
thermalization/isotropization transition in the Bjorken sense
\cite{Bjorken:1982qr,Kovchegov:2005ss}.

The actual calculations are performed easily. Plugging \eq{tzero} into
\eq{h0int} yields
\begin{align}
  h_0 (x^+, x^-) \, = \, 8 \, \Lambda_1^2 \, \Lambda_2^2 \, \theta
  (x^-) \, \theta (x^+).
\end{align}
Using Eqs. (\ref{h1int}), (\ref{hsol}), (\ref{gsol}), (\ref{fsol}),
(\ref{lam}), (\ref{fdsol}), (\ref{lam2}), and (\ref{metric2}) we find
the second order correction to the metric (\ref{2nuc2})
\begin{align}\label{metric2z}
  & g_{--}^{(2)} = \frac{L^2}{z^2} \, \Lambda_1^2 \, \Lambda_2^2 \,
  \left[ - 8 \, \delta (x^-) \, x^+ \, \theta (x^+) \, z^4 -
    \frac{4}{3} \, \delta' (x^-) \, \theta (x^+) \, z^6 - \frac{1}{12}
    \, \delta'' (x^-) \, \delta (x^+) \,  z^8 \right], \notag \\
  & g_{++}^{(2)} = \frac{L^2}{z^2} \, \Lambda_1^2 \, \Lambda_2^2 \,
  \left[ - 8 \, x^- \, \theta (x^-) \, \delta (x^+) \, z^4 -
    \frac{4}{3} \, \theta (x^-) \, \delta' (x^+) \, z^6 - \frac{1}{12}
    \, \delta (x^-) \, \delta'' (x^+) \, z^8 \right], \notag \\
  & g_{+-}^{(2)} = - \frac{1}{2} \, \frac{L^2}{z^2} \, \Lambda_1^2 \,
  \Lambda_2^2 \, \left[ - 16 \, \theta (x^-) \, \theta (x^+) \, z^4 -
    \frac{8}{3} \, \delta (x^-) \, \delta (x^+) \, z^6 + \frac{2}{3}
    \, \delta' (x^-) \, \delta' (x^+) \, z^8 \right], \notag \\ &
  g_{ij}^{(2)} = \frac{L^2}{z^2} \, \delta_{ij} \, \Lambda_1^2 \,
  \Lambda_2^2 \, \left[ 8 \, \theta (x^-) \, \theta (x^+) \, z^4 +
    \frac{4}{3} \, \delta (x^-) \, \delta (x^+) \, z^6 \right],
\end{align}
where the double prime denotes the second derivative. 

One might be concerned with the fact that order-$z^4$ components of
$g_{--}^{(2)}$ and $g_{++}^{(2)}$ contain a negative contributions
localized to the light cone and growing with time. They may be
interpreted, just like \eq{stop} above, as a signal of stopping of the
shock waves. However, our shock waves start out carrying negative
energy and momentum densities. Hence the concept of stopping is
ill-defined for our shock waves, because they themselves are
ill-defined as physical objects, and are only used as some sources
providing us with realistic dynamics of the produced medium in the
forward light cone.

Using Eqs. (\ref{Tsol}) and (\ref{tmngen}) along with \eq{holo} we can
read off the energy density and the pressure components in the forward
light cone from \eq{metric2z}
\begin{align}\label{epp3}
  & \epsilon (\tau) \, = \, \frac{N_c^2}{\pi^2} \, 4 \, \Lambda_1^2 \,
  \Lambda_2^2, \notag \\ & p (\tau) \, = \, \frac{N_c^2}{\pi^2} \, 4
  \, \Lambda_1^2 \, \Lambda_2^2, \notag \\ & p_3 (\tau) \, = \, -
  \frac{N_c^2}{\pi^2} \, 4 \, \Lambda_1^2 \, \Lambda_2^2.
\end{align}
Once again, just like in CGC \cite{Lappi:2003bi,Fukushima:2007ja} and
as was obtained in \cite{Kovchegov:2007pq}, the strongly-coupled
medium starts out very anisotropic, with a negative longitudinal
pressure. Eqs. (\ref{metric2z}), combined with the lower order metric
(\ref{2nuc2}) and Eqs. (\ref{tzero}), allow for a systematic
construction of the metric as an expansion in graviton exchanges, to
be performed elsewhere \cite{future}.

Negativity of the longitudinal pressure $p_3$ in \eq{epp3} is
intimately connected with energy conservation. One can easily see from
\eq{bjhyd} that if the energy density scales as $\epsilon \sim 1/\tau$
then $p_3 =0$. For the energy density which falls off slower with
$\tau$, using \eq{bjhyd} one gets negative $p_3$, in agreement with
\eq{epp3}. For the energy density falling off with $\tau$ faster than
$1/\tau$ one would get positive $p_3$: however, such behavior of
energy density at early time would violate energy conservation. The
net energy of the produced medium at early times is proportional to $E
\sim \epsilon \, \tau$. The energy density which increases faster than
$1/\tau$ at small $\tau$ would then lead to infinite energy of the
produced medium at very early times, violating energy conservation.
Hence energy density at early times can not scale faster than
$1/\tau$, leading to negative or zero longitudinal pressure $p_3$.


\section{The Dilaton}
\label{dilaton}

Let us briefly touch upon one related topic which may become important
at higher order in graviton exchanges. Immediately after the collision
of two heavy ions the medium is not equilibrated yet. The magnitudes
squared for the chromo-electric and chromo-magnetic fields are not
equal to each other. This means that the expectation value of the
gluonic field strength squared should not be zero. In fact, at weak
coupling a CGC calculation at the lowest non-trivial order (order
$\as^3$) performed along the lines of
\cite{Kovchegov:2005ss,Kovchegov:1997ke} yields
\begin{align}\label{trF}
  \langle \mbox{tr} F_{\mu\nu}^2 \rangle \, = \, - 4 \, \as^3 \, C_F
  \, \frac{A^2}{S^2_\perp} \, \int \frac{d^2 k_T}{k_T^2} \, \left[ J_0^2
    (k_T \tau) - J_1^2 (k_T \tau) \right].
\end{align}
Here $A$ and $S_\perp$ are the atomic number and the transverse area
of the two identical colliding nuclei and $C_F = (N_c^2 -1)/ 2 N_c$.
To obtain \eq{trF} one should substitute the gluon field from Eq. (12)
in \cite{Kovchegov:1997ke} into the Abelian part of $\mbox{tr}
F_{\mu\nu}^2$ and average the resulting expression in the wave
functions of both nuclei (see also
\cite{Kharzeev:2001vs,Kovchegov:2005ss}).  At early times \eq{trF}
gives
\begin{align}\label{trFt}
  \langle \mbox{tr} F_{\mu\nu}^2 \rangle \bigg|_{Q_s \tau \ll 1} \,
  \approx \,- 4 \, \pi \, \as^3 \, C_F \, \frac{A^2}{S_\perp^2} \, \ln
  \frac{1}{Q_s^2 \, \tau^2}
\end{align}
where $Q_s^2 = 4 \, \pi \, \as^2 \, A/S_\perp$ is the saturation scale
of the nuclei which regulates the infrared divergence in \eq{trF} when
higher order rescatterings are included.  Similar to
\cite{Lappi:2003bi,Fukushima:2007ja} one may conclude that the scaling
(\ref{trFt}) is true for the all-order classical gluon field
\cite{Kovner:1995ts,Kovner:1995ja,Kovchegov:1997ke,Gyulassy:1997vt,Krasnitz:2003nv,Krasnitz:2003jw,Krasnitz:1999wc,Krasnitz:2002mn,Kovchegov:2000hz}.
We get
\begin{align}\label{logdiv}
  \langle \mbox{tr} F_{\mu\nu}^2 \rangle \bigg|_{Q_s \tau \ll 1} \,
  \propto \, - \frac{Q_s^4}{\as} \, \ln \frac{1}{Q_s^2 \, \tau^2}.
\end{align}
As $\langle \mbox{tr} F_{\mu\nu}^2 \rangle = 2 (B^2 - E^2)$ with $B$
and $E$ the chromo-magnetic and chromo-electric fields, we conclude
that at weak coupling the medium at the early stages of the collisions
is dominated by chromo-electric fields (see also \cite{Lappi:2006fp}).

To introduce non-zero expectation value for $\mbox{tr} F_{\mu\nu}^2$
in AdS one needs to include the dilaton field $\varphi$, as
\cite{Klebanov:2000me,Balasubramanian:1998de,Heller:2007qt}
\begin{align}\label{dila}
  \frac{1}{4 \, g_{YM}^2} \, \langle \mbox{tr} F_{\mu\nu}^2 \rangle \,
  = \, \frac{N_c^2}{2 \, \pi^2} \, \lim_{z \rightarrow 0}
  \frac{\varphi (x^+, x^-, z)}{z^4}.
\end{align}
The dilaton couples to the metric through modified Einstein equations
(\ref{ein4}) and through the Klein-Gordon equation (\ref{KG}).

The metrics of the incoming shock waves like (\ref{1nuc}) are
solutions of Einstein equations (\ref{ein2}) with zero dilaton fields.
This implies that the dilaton field is zero for a single nucleus and,
using \eq{dila}, $\langle \mbox{tr} F_{\mu\nu}^2 \rangle =0$ for a
single nucleus as well. This agrees with the fact that at weak
coupling $\langle \mbox{tr} F_{\mu\nu}^2 \rangle =0$ too, as the
electric and magnetic fields of a single ultrarelativistic charge are
equal to each other (the equivalent photon/gluon approximation)
\cite{Kovchegov:1996ty,Jalilian-Marian:1997xn}.

At strong coupling the dilaton field may become non-zero after the
collision. As one can see from \cite{Heller:2007qt}, an investigation
of late-time dynamics of the strongly-coupled medium requires a
dilaton field leading to non-zero $\langle \mbox{tr} F_{\mu\nu}^2
\rangle$. Since the dilaton field of each of the shock waves is zero,
we may only expect that the dilaton field produced in the collision
would depend on energy-momentum tensors of both shock waves.
Therefore, using the expansion in $\Lambda_1^2$ and $\Lambda_2^2$ of
Sect. \ref{our_shock}, one may expect that at the lowest non-trivial
order
\begin{align}\label{F}
  \langle \mbox{tr} F_{\mu\nu}^2 \rangle \, \sim \, \Lambda_1^2 \,
  \Lambda_2^2
\end{align}
corresponding to the dilaton field (see \eq{dila})
\begin{align}\label{dile}
  \varphi (x^+, x^-, z) \, \sim \, \Lambda_1^2 \, \Lambda_2^2 \, z^4.
\end{align}
However, the sign of the dilaton field and $\langle \mbox{tr}
F_{\mu\nu}^2 \rangle$ can not be determined from these dimensional
considerations. (In fact Eqs. (\ref{ein4}) and (\ref{KG}) are
invariant under $\varphi \rightarrow - \varphi$ and do not fix the
sign of the dilaton field.) Instead of the logarithmic divergence
(\ref{logdiv}) of the perturbation theory, we anticipate the
expectation value of $\langle \mbox{tr} F_{\mu\nu}^2 \rangle$ to go to
a constant at early times.

The dilaton field from \eq{dile} would affect the metric only at the
order $\Lambda_1^4 \, \Lambda_2^4$, as can be seen from \eq{ein4}, and
therefore we could safely neglect it in the above discussion and in
\cite{Kovchegov:2007pq}. It would only enter \eq{ein4} at the same
order as the four-graviton exchanges.

We stress that while we do not know whether non-zero dilaton field
would arise at higher orders in our expansion in graviton exchanges,
if it does come in we expect it to be of the form shown in \eq{dile}
(at the lowest order) leading to a constant $\langle \mbox{tr}
F_{\mu\nu}^2 \rangle$ at early times shown in \eq{F}.


\section{Conclusions}
\label{conc}

Let us summarize the main points of our paper. In Sect.
\ref{real_shock} we have demonstrated that a collision of real and
physical shock waves in AdS would lead to a full stopping of the shock
waves. It is likely that a black hole would be created in AdS space
immediately afterwards. For the gauge theory this implies that two
nuclei colliding at very strong coupling would stop almost immediately
after the collision, after a time interval of the order of $t_{stop}
\sim 1/\Lambda$ with $\Lambda$ some typical transverse momentum of the
problem. Thus for RHIC and LHC one might expect $t_{stop} \approx
1$~fm. After the stop, creation of the black hole in the bulk likely
translates into thermalization and Landau hydrodynamics description of
the dynamics for the created medium. This is a possible scenario
advocated in \cite{Steinberg:2004vy}.

However, we consider it to be more likely that the physics of the
initial stages of heavy ion collisions is weakly coupled. This point
of view is supported by the many successes of CGC approaches to heavy
ion collisions \cite{Kharzeev:2000ph,Kharzeev:2001yq}. In the weak
coupling scenario the hard (large-$x$) parts of the nuclear wave
functions simply go through each other in the collisions without
deflection or recoil. At the same time the produced thermalized medium
should still be strongly-coupled.  While no comprehensive single
description of both the weakly-coupled early stages and the
strongly-coupled medium in the final state exists, in Sect.
\ref{our_shock} we constructed a model which appears to capture the
main features of the weakly-coupled initial state by allowing the
energy density to be negative (only) on the light cone. The
corresponding shock waves carry both positive and negative energy
density and are thus unphysical. They need to be thought about as some
external sources for the gravitational field used in gauge-gravity
duality. The energy density of the produced medium in the forward
light-cone is non-negative: in fact we recovered our earlier result of
\cite{Kovchegov:2007pq} that the energy density starts out as a
constant at early proper times.

We have thus arrived at the following conclusion. If the coupling
constant in heavy ion collisions is large throughout the collision
this would lead to nuclear stopping followed by Landau hydrodynamics.
In the more realistic scenario, according to present phenomenological
evidence, in which both the nuclear wave functions and the primary
particle production are weakly-coupled, Bjorken hydrodynamics could be
still achieved if the coupling constant quickly becomes large. Indeed,
as shown in \cite{Janik:2005zt}, a purely strong coupling approach to
the study of late-time dynamics leads to Bjorken hydrodynamics.
However, as we argued in the paper, Bjorken hydrodynamics can not
result from strong coupling dynamics only. \\

When this paper was in the final stages of preparation, a preprint
\cite{Gubser:2008pc} was posted on the arXiv, where a similar
conclusion about stopping of shock waves has been reached. Also, very
recently a new version of \cite{Grumiller:2008va} appeared on the
arXiv, where the possibility of the shock wave energy-momentum tensor
as in our \eq{Tzero} was briefly mentioned.


\acknowledgments

We would like to thank Samir Mathur for many highly educational and
informative discussions. We thank Ulrich Heinz, Al Mueller and Krishna
Rajagopal for encouraging discussions.

This research is sponsored in part by the U.S. Department of Energy
under Grant No. DE-FG02-05ER41377.


\appendix

\renewcommand{\theequation}{A\arabic{equation}}
  \setcounter{equation}{0}
\section{A model for the energy-momentum tensor of an ultra-relativistic nucleus}
\label{A}

We begin by considering the classical electromagnetic potential of a point-like
particle of charge $g$ moving at speed $v$ along the positive $z$ direction.
In the covariant gauge it is given by
\begin{align}
  A_{\pm}=\frac{g}{4\,\pi}\frac{(1\mp
    v)/\sqrt{2}}{\left[\frac{1}{2}\left((1+v) \,
        x^--(1-v)\,x^{+}\right)^2+(1-v^2)\,x_{\perp}^2\right]^{1/2}}\,,\quad
  A_i=0\,\, \ (i=1,2).
\end{align}

The energy-momentum tensor associated to this field is
\begin{align}
  T_{\mu\nu}=-F_{\mu\rho}\,F_{\nu}^{\,\,\rho}+\frac{1}{4}\,\eta_{\mu\nu}\,F_{\rho
    \, \sigma} F^{\rho \, \sigma} \,,
\end{align}
where $\eta$ is the Minkowski metric in four dimensions. In the limit $v\to
1$, its only non-vanishing component is 
\begin{align}\label{tpp}
  T_{--}=\left(\partial_i\,A_-\right)^2=\frac{\alpha}{2\pi}
  \frac{(1-v^2)^2\,x_{\perp}^2}{\left[2x^{-2}+(1-v^2)x_{\perp}^2\right]^{3}}\,,
\end{align}
where $\alpha=g^2/4\pi$. In the strict limit $v\!=\!1$, $T_{--}$ is
singular at $x^-\!=\!0$. To clarify the nature of this singularity we
consider the following integral:
\begin{align}\label{eint}
  \int^{\infty}_{-\infty}\!dx\,\frac{\epsilon^4}{\left(x^2+\epsilon^2\right)^3}\,f(x)\,
  =(2\pi\, i)\,
  \epsilon^4\,\frac{1}{2!}\,\frac{d^{\,2}}{d\,z^2}\left. \frac{f(z)}{(z+i\, |\epsilon|)^3}\right\vert_{z=i\, |\epsilon|}
  \stackrel{\epsilon\rightarrow
    0}{=}\frac{3\,\pi}{8}\, \frac{f(0)}{|\epsilon|},
\end{align}
where $f(x)$ is an arbitrary analytic function that falls off at
infinity rapidly enough for the previous integral to be well defined.
From \eq{tpp} and \eq{eint} we get
\begin{align}
  T_{--}=\frac{\alpha}{2\pi}\frac{8}{3\pi} \,
  \frac{\delta(x^-)}{|x^-|} \, \frac{1}{x_{\perp}^2}
\end{align}

The previous equation serves as a starting point to build up a model
for the energy-momentum tensor of an ultra-relativistic nucleus, $
T_{--}^{\,nucl}$. We envisage the nucleus as consisting of $A$ {\it
  nucleons}, each of them containing $N_c^2$ {\it{valence}} gluons.
Assuming an uniform distribution of nucleons inside the nucleus,
averaging over transverse position and summing over all nucleons, we
write
\begin{align}\label{taver}
  \langle T_{--}^{\,nucl}\rangle=
  N_c^2\,\frac{A}{S_{\perp}}\int\,d^2x_{\perp}\,
  T_{--}\,=\frac{N_c^2\,\alpha \, N_c}{\pi} \,
  \frac{\delta(x^-)}{|x^-|} \,
  \frac{8}{3}\frac{A}{S_{\perp}}\,\ln\left(\frac{R}{\rho}\right),
\end{align}
where $S_{\perp}$ is the nuclear transverse area, $R$ is the nuclear
radius and $\rho$ is an UV cutoff introduced to regulate the singular
behavior at $x_{\perp}\!=\!0$. The extra factor of $N_c$ in
\eq{taver} comes from calculating the color factor for non-Abelian
energy-momentum tensor at the lowest order in the coupling. For
simplicity, we consider a cylindrical nucleus such that
$S_{\perp}\approx \pi\,A^{2/3}\,R_N^{2/3}$, with $R_N$ the nucleon's
radius. Introducing the dimensionful scale
\begin{align}
  \Lambda^2 \, \equiv \,
  \frac{4}{3\,\pi^3 \, R_N^2} \, \ln\left(\frac{R}{\rho}\right)\,,
\end{align}
we rewrite \eq{taver} as
\begin{align}\label{tmn1}
  \langle T_{--}^{\,nucl}\rangle= N_c^2\, \alpha\, N_c \, 4 \, \pi \,
  \Lambda^2\,A^{1/3}\, \frac{\delta(x^-)}{2 \, |x^-|}.
\end{align}
Defining 't Hooft coupling $\lambda$ by \eq{thooft} we rewrite
\eq{tmn1} as
\begin{align}\label{tmn2}
  \langle T_{--}^{\,nucl}\rangle= N_c^2\,\lambda \,
  \Lambda^2\,A^{1/3}\, \frac{\delta(x^-)}{2 \, |x^-|},
\end{align}
which is now ready to be cast in the form of \eq{sea} at strong 't
Hooft coupling by replacing $\lambda \rightarrow \sqrt{\lambda}$. The
singularity in $\delta(x^-) / |x^-|$ can be regularized by replacing
$1/|x^-|$ with the light-cone momentum of the valence gluon $p^+$,
which would reduce \eq{tmn2} to \eq{Tmu} with $\mu$ given by \eq{mu}.
The energy-momentum tensor in \eq{tmn2} may serve as the source for
the one-graviton exchange shock wave metric in \eq{1nuc}.


\renewcommand{\theequation}{B\arabic{equation}}
  \setcounter{equation}{0}
\section{Solution of Equations (\protect\ref{ein})}
\label{B}
Here we complete the solutions of the Einstein equations for the
second order correction to the metric, $g_{\mu\nu}^{(2)}$. Analogously
to what we did for $h$ in \eq{hsol_long}, we write the unknown
functions $g$, $f$ and $\tilde f$ in \eq{metric2} in the form of a
power series in $z^2$, starting at order $z^4$:
\begin{align}\label{gser}
  g (x^+, x^-, z) \, = \, z^4 \, \sum\limits_{n=0}^\infty g_n (x^+,
  x^-) \, z^{2n} \notag \\ f (x^+, x^-, z) \, = \, z^4 \,
  \sum\limits_{n=0}^\infty f_n (x^+, x^-) \, z^{2n} \notag \\
  {\tilde f} (x^+, x^-, z) \, = \, z^4 \,
  \sum\limits_{n=0}^\infty {\tilde f}_n (x^+, x^-) \, z^{2n} \, .
\end{align}
Inserting these expansions into Eqs. (\ref{zz}), (\ref{-z}) and
(\ref{+z}) we find
\begin{align}
  g_n+2h_n=\frac{2}{3}\,\delta_{n,2}\,t_1(x^-)\,t_2(x^+)\,,\\
  f_{n,x^-}+h_{n,x^+}=-\frac{1}{12}\,\delta_{n,2}\,t_1(x^-)\,t'_{2}(x^+)\,,\\
  {\tilde
    f}_{n,x^+}+h_{n,x^-}=-\frac{1}{12}\,\delta_{n,2}\,t'_1(x^-)\,t_{2}(x^+)\,,
\end{align}
which straightforwardly lead to the following relation between the
metric coefficients
\begin{align}
  g(x^+,x^-,z)=-2\,h(x^+,x^-,z)+\frac{2
  }{3}\,z^8\,t_1(x^-)\,t_2(x^+)\,,\label{solg} \\
  f (x^+,x^-,z)=-\frac{\partial_-}{\partial_+}\left(h(x^+,x^-,z)+
    \frac{1}{12}\,z^8\,t_1(x^-) \, t_2(x^+)\right)\,,\label{solf1}  \\
  {\tilde f}
  (x^+,x^-,z)=-\frac{\partial_+}{\partial_-}\left(h(x^+,x^-,z)+
    \frac{1}{12}\,z^8\,t_1(x^-)\, t_2(x^+)\right). \label{solf2}
\end{align}

To complete the solution of Einstein equations Eqs. (\ref{ein}) we
insert the relations in Eqs. (\ref{solg}-\ref{solf2}) into the
remaining Einstein equations, Eqs. (\ref{--}-\ref{+-}), which we have
not used yet. We find that, while Eqs. (\ref{--}) and (\ref{++}) are
trivially satisfied, \eq{+-} provides the following equation for
$h(x^+,x^-,z)$:
\begin{align}\label{B8}
3\,h_z-z\,h_{zz}+2\,z\,h_{x^+x^-}=\frac{8}{3}\,z^7\,t_1(x^-)\,t_2(x^+)\,,
\end{align}
which can be easily reduced to \eq{heq} with the general solution
given in \eq{hsol_long}. Finally, in order to complete the solution of
the Einstein equations (\ref{ein}) we must determine the unknown
coefficient $h_0(x^+,x^-)$ in \eq{hsol_long}. We do so by imposing
causality of the solution which, as discussed in Sect. \ref{gensol},
implies that all the second order corrections to the metric in
\eq{metric2} must be zero before the collision. Given the relation
between the metric coefficients in Eqs.  (\ref{solg}-\ref{solf2}), it
is sufficient to require causality of $h(x^+,x^-,z)$. We first note
that by requiring that
 \begin{align}\label{B9}
  (\partial_+ \, \partial_-)^2 \, h_{0} (x^+, x^-) \, = \, 8 \, t_1
  (x^-) \, t_2 (x^+)
\end{align}
we satisfy the initial ($h=0$ before the collision) and boundary
($h(x^+, x^-,z=0)=0$) conditions on the solution of \eq{B8}. \eq{B9}
is exactly \eq{h0eq} above.  This condition makes the infinite series
in Eqs.  (\ref{hsol_long}) and (\ref{gser}) terminate, as previously
announced, and satisfies the causal initial conditions. As the
solution of \eq{B8} for a given set of initial and boundary conditions
is unique, we conclude that the condition (\ref{B9}) singles out the
only causal solution of Einstein equations. (This implies that
solutions obtained from \eq{hsol_long} without terminating the
infinite series are not causal.) This proves that \eq{B9} gives us the
only solution of \eq{B8} satisfying the needed initial and boundary
conditions.  The complete solution of Einstein equations (\ref{ein})
is given by Eqs.  (\ref{hsol}), (\ref{gsol}), (\ref{fsol}) and
(\ref{fdsol}).



\providecommand{\href}[2]{#2}\begingroup\raggedright\endgroup


\end{document}